\documentclass[structabstract]{aa}  
\usepackage{graphicx,subfig}
\usepackage{natbib}
\usepackage{longtable}
\usepackage[english]{babel}
\usepackage{txfonts}
\def \xmmn {\emph{XMM-Newton }}
\def \xmm {\emph{XMM-Newton}}
\def \chandra {\emph{Chandra }}

\begin{document}

\title{Cool core remnants in galaxy clusters}

\author{Mariachiara Rossetti
         \inst{1}
                  \and
         Silvano Molendi
         \inst{1}
         }

\institute{Istituto di Astrofisica Spaziale e Fisica Cosmica, INAF, via Bassini 15, I-20133 Milano, Italy}


\abstract
   {X ray clusters are conventionally divided into two classes: ``cool core'' (CC) and ``non cool core'' (NCC) objects, on the basis of the observational properties of their central regions. Recent results have shown that the cluster population is bimodal (Cavagnolo et al. 2009).}
   {We want to understand whether the observed distribution of clusters is due to a primordial division into two distinct classes rather than to differences in how these systems evolve across cosmic time.}
   {We systematically  search the ICM of NCC clusters in a subsample of the B55 flux limited sample of clusters for regions which have some characteristics typical of cool cores, namely low entropy gas and high metal abundance}
   {We find that most NCC clusters in our sample host regions reminiscent of CC, i.\, e.\, characterized by relative low entropy gas (albeit not as low as in CC systems) and a metal abundance excess. We have dubbed these structures ``cool core remnants'', since we interpret them as what remains of a cool core after a heating event (AGN giant outbursts in a few cases and more commonly mergers). 
We infer that most NCC clusters have undergone a cool core phase during their life.
The fact that most cool core remnants are found in dynamically active objects provides strong support to  scenarios where cluster core properties are not fixed ``ab initio'' but evolve across cosmic time.}
   {}


\maketitle


\section{Introduction}
Galaxy clusters are often divided by X-ray astronomers into two classes: ``cool core''(CC) and ``non-cool core'' (NCC) clusters. The former are characterized by a set of properties: a prominent surface brightness peak, usually roughly coincident with the center of the large scale X-ray isophotes and with the position of the brightest central galaxy (BCG), associated to a decrease of the temperature profile in the inner regions and to a positive gradient of the metal abundance profile. In the central regions of these clusters, the cooling time is significantly smaller than the Hubble time but the radiative losses are balanced by some form of heating, that in the currently prevailing scenario is attributed to the central AGN, which prevents the gas from cooling indefinitely and  flowing inward. NCC clusters are characterized by the lack of these observational features. Several indicators based on these observational characteristics have been proposed to classify clusters into these categories: e.\,g.\,
temperature drop (\citealt{sanderson06, sanderson09}), cooling time \citep{bauer05}, a composition of these two criteria (\citealt{dunn05, dunn08}), the slope of the gas density profile at a given radius \citep{vikh_garching} and core entropy \citep{cava09}. 
In a recent paper (\citealt{pap1}, hereafter Paper I), we suggested a robust indicator to classify clusters (see also Sect.\, \ref{sec_classif}), based on pseudo-entropy gradients. The classification based on this indicator improves the traditional classification scheme based on the temperature drop and is essentially equivalent to classifications based on the cooling time.
Increasing attention has been devoted to the statistics of CC, both in the local Universe and at higher redshifts. While the precise results depend strongly on the indicator used to classify clusters, the fraction of CC is considered to be about $50\%$ of the clusters population. There are some indication that  the cluster population is bimodal \citep{cava09} but there are also some intermediate objects which are not easily classified (Paper I).\\
One of the open questions in the study of galaxy clusters, concerns the origin of this distribution. 
The original model which prevailed for a long time assumed that the CC state was a sort of ``natural state'' for the clusters and the observational features were explained within the context of the old ``cooling flow'' model: radiation losses cause the gas in the centers of these clusters to cool and to flow inward. Clusters were supposed to live in this state until disturbed by a ``merger''. Indeed, mergers are very energetic events that can shock heat \citep{burns97} and mix the ICM \citep{gomez02, ritchie02}: through these processes they were supposed to efficiently destroy cooling flows. After the mergers, clusters were supposed to relax and go back to the cooling flow state in a sort of cyclical evolution.
With the fall of the ``cooling flow'' brought about by the \xmmn and \chandra observations (e.\,g.\, \citealt{peterson01} and \citealt{molepizzo01}), doubts were cast also on the interpretation of mergers as the dominant mechanism which could transform CC clusters into NCC. More generally speaking, the question arose whether the observed distribution of clusters was due to a primordial division into the two classes rather than to evolutionary differences during the history of the clusters.  \\
\citet{mccarthy04, mccarthy08} noticed the absence of systems that resemble observed NCC clusters in cosmological simulations, suggesting that mergers could not be the origin of the cluster distribution. They proposed that early episodes of non gravitational pre-heating may  explain the dichotomy; they envisage a scenario in which NCC clusters have been pre-heated to levels greater than $\sim 300\, \rm{keV}\, \rm{cm}^{2}$ and do not have enough time to develop a cool core while CC clusters have been pre-heated to lower levels and need an additional source of present-day heating to offset cooling. With this model, \citet{mccarthy08} explained observed entropy and gas density \chandra profiles for CC and \emph{ROSAT} gas density profiles for NCC. 
\citet{ohara06} supported the primordial model showing that the scatter in scaling relations is larger for CC clusters than for NCC, suggesting that CC are not more relaxed than NCC systems. Moreover the  two body idealized simulations by \citet{poole08} showed that mergers cannot produce extended ``warm'' cores and cannot destroy metal abundance gradients. \\
However, the evolutionary ``merger'' scenario has been continuously supported by observations.
For instance, \citet{sanderson06} compared temperature and cooling time profiles of a sample of clusters observed by \chandra with indicators of merger activity and suggested that merger may be the primary factor in preventing the formation of CC. More recently, \citet{sanderson09b} have shown in a large sample of objects that the X-ray/BCG projected offset correlates with the gas density profile, which can be considered an indicator of the CC state. If the X-ray/BCG offset ``measures'' the dynamical state of the cluster, this result implies that the cool core strengths diminishes in more dynamically disturbed clusters. 
Another indicator of dynamical activity is the presence of extended radio halos on Mpc scales, whose origin is likely related to turbulent acceleration driven by mergers (\citealt{ferrari08} for a recent review). None of the clusters which are found to host a radio halo (on Mpc scales) in a complete survey \citep{venturi08, brunetti09} can be classified as a CC object. These relations are indeed expected if mergers can efficiently destroy cool cores.\\
An interesting model has been proposed by \citet{motl04}, who suggested a double role for mergers in the cluster formation process: while they can shock heat the ICM they also efficiently mix the ICM and participate to the formation of CC by providing cool gas. By analyzing simulated temperature and surface brightness maps, they find that observational signatures of a cool core may disappear after mergers but cool gas remains in the systems at all time. Indeed mergers are complicated phenomena during which the ICM of the interacting objects is mixed. Also  \citet{ascas06}, with different simulations with the purpose of studying the formation of cold fronts, showed that an interacting subcluster may donate its cool gas to the main cluster and that, during the merging processes, many hydrodynamic effects contribute to the mixing of the gas. \\
Recently, we found an unexpected and interesting result from the analysis of the metalicity profiles in a large sample of clusters (Paper I): some clusters that cannot be classified as CC show a metal abundance excess at their center, witout a significant temperature decline or an X-ray brightness excess.
 We suggested that at least some NCC clusters have spent part of their lives as CC objects and therefore there cannot be a complete primordial separation of the two classes of objects.\\
However, further analysis is needed on this result to use it to gain insight on the origin of the distribution of CC-NCC objects. We would like to better characterize these regions and to know if and to what extent they are common in the population of NCC objects.
The results in Paper I, as well as those described above comparing observations and simulation, are based on  ``uni-dimensional'' properties (entropy and metal abundance profiles). Indeed, global properties and unidimensional profiles are easy to derive and it is possible to compare them quantitatively but they inevitably result in a loss of information. There is now a large number of thermodynamic maps in the literature, which show that spherical symmetry is not generally fulfilled in clusters and revealed the presence of regions with ``anomalous'' characteristics outside of the cores (e.g \citealt{sun02,rossetti06} and many others ). \\
In this paper, we present a systematic two-dimensional analysis on a sample of 35 clusters observed with \xmm,  to look for and characterize regions with a significant metal excess in NCC clusters, as those found in Paper I. The outline of the paper is as follows: in Sect.\,2 we describe the sample and the data analysis, in Sect.\,3 we discuss the classification schemes and we define ``CC remnants'' in Sect.\,4.  We interpret our results in Sect.\,5 and  we summarize our findings in Sect.\,6. Quoted confidence intervals are 68\% for one interesting parameter unless otherwise stated. All results are given assuming a $\Lambda$CDM cosmology with $\Omega_{\rm{m}} = 0.3$, $\Omega_\Lambda = 0.7$, and $H_0 = 70$ km s$^{-1}$ Mpc$^{-1}$.


\section{Data analysis}

\subsection{The sample}
The starting point of the sample of galaxy clusters described in this paper is the ``B55'' X-ray flux limited sample \citep{edge90}. In order to have a significant coverage of the cluster within EPIC field of view, we have eliminated the nearest clusters and considered only those with redshift $z>0.03$. Then we analyzed all public \xmmn observations, including mosaics and multiple observations, as described in Sect. \ref{datareduction}. We discarded observations where, after soft proton cleaning, the effective exposure time ($\rm{MOS1} + \rm{MOS2} + \rm{pn}$) is lower than 25 ks, except in the case of mosaics where observations with $t_{exp}<25$ ks have been used for images but not for spectral analysis.\\
The list of the clusters which survived our selection criteria is given in Table \ref{tablist}.  A644, A2244 have not been observed by \xmm, while all the observations of A3391, A1736, A2142, A2063 and A1651 are badly contaminated by soft protons ($t_{exp}<25$ ks) and have been discarded. 
The clusters Cygnus A has been discarded in a second phase of the analysis because of the presence in the core of the radio galaxy QSO B$1957+405$, featuring large hotspots well detected in X-rays, which cannot be easily subtracted from the core spectrum.  \\
\addtocounter{table}{1}

\subsection{General data reduction}
\label{datareduction}
We retrieve Observation Data Files (ODF) from the \xmmn archive and process them with SAS software version $7.0$. The event files produced by this standard analysis technique are then cleaned to remove soft proton flares with a double filtering process. As a first step, we produce the light curve in a hard energy band (10-12 keV) and we remove all time periods with count rates exceeding a fixed threshold $0.025\, \rm{cts/s}$ for the MOS detectors and  $0.050\, \rm{cts/s}$ for the pn. This procedure allows to remove most flares, but softer flares may survive. In the second step, we apply a $\sigma$ clipping technique to the histogram obtained from the light curve in $2-5$ keV energy range. 
For each observation, we have calculated the ``in over out ratio'', $\rm{R}_{\rm{SB}}$ \citep{deluca03}, to identify observations badly contaminated by a quiescent soft proton component. As outlined in Paper I, the ratio has been calculated in an external annulus at $E>9$ keV to reduce the contribution of the cluster emission which fills all the FOV, especially for the nearest objects. We have discarded only one observation (and therefore the cluster A2065) where the $\rm{R}_{\rm{SB}}$ of the two MOS is larger than $2.0$. 
After soft proton cleaning we filter the event files, according to pattern and flag criteria, and we remove by eye brightest point sources.\\   
Since we are mainly interested in characterizing the thermodynamic properties of the ICM in the central and brightest regions of the clusters of our sample, advanced procedures to treat the background \citep{leccardi08a} are not strictly necessary. This is also the reason why we could discard only badly contaminated observations with $\rm{R}_{\rm{SB}}>2.0$.  
As background event files, we merged nine ``blank sky'' field observations, as commonly done, and we extract images and  spectra for the background as for the source observations. We calculate a normalization factor $Q$ for each cluster observation, to take into account possible temporal variations of the instrumental background. The  normalization factor is the count rate ratio between source and background observations in an external ring ($10^\prime-12^\prime$) beyond 9 keV. Background images and spectra are scaled by $Q$ before subtraction from source images and spectra.

\subsection{Two-dimensional analysis}
\label{2D}
For all the clusters in our sample we prepared two dimensional maps of the main thermodynamic quantities, starting from EPIC images. To do this we have used a modified version of the adaptive binning + broad band fitting technique described in \citet{rossetti06}, where we have substituted the \citet{cappellari}  adaptive binning algorithm with the weighted Voronoi tessellation by \citealt{diehl06} \citep{rossettith}. \\
Especially in clusters undergoing major mergers, where there is no spherical symmetry, maps are a fundamental tool to select interesting regions for a proper spectral analysis, which is necessary to complement the thermodynamical information with chemical information. Indeed a ``blind'' spectral analysis in concentric annuli would not allow to detect interesting features in our clusters.\\

\subsection{One-dimensional analysis}
\subsubsection{Profiles}
\label{prof}
As a first step, we have performed spectral extraction and analysis in concentric annuli. As discussed in Sect. \ref{2D}, the main drawback of this technique is that the assumption of spherical symmetry is not always fulfilled in our clusters and in some cases even the choice of the center can have a strong impact on the observed properties. Therefore we have used the thermodynamic maps (Sect.\, \ref{2D}) to identify those clusters where the deviations from the spherical symmetry are larger and we have decided not to perform radial analysis in four well known merging objects: A3667 \citep{briel04}, A2256 \citep{sun02}, A754  \citep{henry04} and A3266  \citep{finog06} \\ 
For the remaining clusters of the sample, where we do not observe large displacements between the surface brightness peak and the entropy minimum, we select as center of symmetry the surface brightness peak, even if the large scale isophotes have another center. This allows to better describe the ICM properties in the more central regions of the clusters. \\
We extract spectra from annular regions around the selected center.
For each instrument (MOS1, MOS2 and pn) and each region we extract source and background spectra and we generate an effective area (ARF) file. Then we associate to the spectrum a redistribution matrix file, appropriated for the instrument and, in the pn case, for the position of the selected region in the detector. \\
We perform spectral fitting, using the XSPEC v11.3 package, for each spectrum separately in the energy range 0.5-10 keV. We use an absorbed mekal model (WABS*MEKAL), where the $N_{H}$ is fixed to the input galactic value\footnote{The only exception is A478, where we found a difference of about a factor of two between the galactic value ($1.48\, 10^{21}\, \rm{cm}^{2}$, Dickey \& Lockman) and the best fit value ($2.7\, 10^{21}\, \rm{cm}^{2}$)} and the redshift is allowed to vary in a small range (width $\simeq 0.02$) around the nominal value. Temperature, metal abundance\footnote{For solar abundances we refer to \citet{anders89}} and normalization are the free parameters of the fit (following the prescription in \citealt{leccardi08b} even negative values are allowed for the metal abundance). Best fit results obtained from the three instruments and from multiple observations of the same region are then averaged together with a weighted mean. This allows to produce projected temperature, metal abundance and surface brightness profiles for each cluster.\\
Finally, we perform deprojection of the profiles, following the technique described in \citet{ettori02}, to derive three-dimensional density and temperature profiles. Combining them, we obtain entropy and cooling time profiles.

\subsubsection{IN and OUT regions}
\label{inandout}
In order to classify clusters according to the scheme described in Paper I, we extracted spectra in a IN region and in a OUT reference region, whose radii are defined as a fixed fraction of $R_{180}$ ($r<0.05\, R_{180}$ and $0.05\, R_{180}<r<0.2\, R_{180}$, respectively).  $R_{180}$ has been calculated as
\begin{equation}
R_{180}=1780 \left(\frac{kT}{5\, \rm{keV}} \right)^{1/2}h(z)^{-1}\, \rm{kpc},
\end{equation}
where $h(z)=(\Omega_m(1+z)^3+\Omega_{\Lambda})^{1/2}$ and $kT$ is the temperature of the cluster \citep{arnaud05, leccardi08a}.
This mean temperature has been first calculated starting from the profiles (eventually excluding the bins of the temperature drop in CC clusters).
After extraction of the OUT spectra, we have re-calculated  $R_{180}$ using as mean temperature $T_{out}$, i.e. the temperature obtained with a spectral fit in the region  $0.05\, R_{180}<r<0.2\, R_{180}$. We have then calculated the differences between  the new estimate of $0.2R_{180}$ and the old one. Once converted in arcseconds, we have compared this difference with the \xmmn point spread function: if the two estimates of $0.2R_{180}$ differed by more than $15\arcsec$, we extracted again spectra with the new $R_{180}$ and iteratively repeated the procedure until differences were smaller than the PSF. 
In the case of the four clusters without radial profiles (A754, A2256, A3266 and A3667), the first estimate of the mean temperatures has been performed directly from the temperature maps. The list of the final $T_{out}$ and $R_{180}$ can be found in Table \ref{clusters}.\\
As discussed in Sect. \ref{2D}, the choice of the center has been performed starting from the thermodynamic maps. More specifically, we have selected as a center the position of the minimum in the pseudo-entropy map. 
In most cases, the position of the entropy minimum coincides with the surface brightness peak, while in other clusters there is a significant displacement, but usually smaller then the radius of the IN regions, with the exception of A3667. Moreover the choice of the position of the entropy minimum as a center instead of the surface brightness peak does not alter significantly the value of the pseudo entropy ratio, except for the case of A3667. This is a well studied merging cluster (\citealt{vikh01b}, \citealt{briel04}): the low-entropy gas is concentrated at the position of a prominent cold front, about 500 kpc SE from the surface brightness peak (see Figure in Appendix B).   


\section{Classification schemes}
\label{sec_classif}

\begin{figure}
   \centering
   \resizebox{\hsize}{!}{\includegraphics{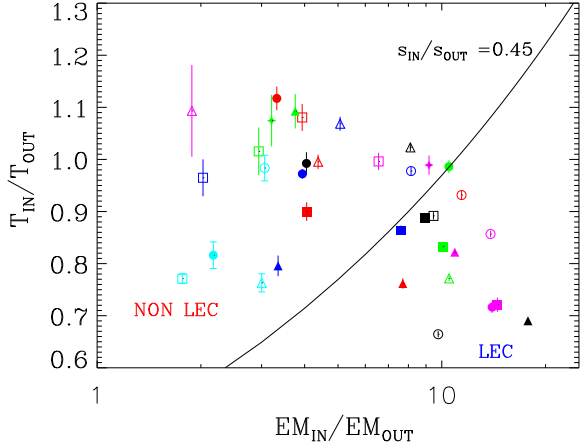}}
   \caption{Comparison of temperature and emission measure ratios for the clusters in our sample. The solid line represent the threshold used to divide clusters between low entropy clusters and non low entropy clusters, corresponding to $s_{IN}/s_{OUT}=0.45$. The symbols and colors are as in Table \ref{tablist}.}
   \label{rat_T_vs_EM}
 \end{figure}
As in Paper I we plot the temperature ratios ($T_{IN}/T_{OUT}$) versus the emission measure ratios ($EM_{IN}/EM_{OUT}$)  for the clusters of our sample (Fig.\,\ref{rat_T_vs_EM}) and we use pseudo-entropy ratios to divide clusters into two classes (Table \ref{clusters}). We recall here the definition of  the pseudo-entropy ratio\footnote{X-ray astronomers usually define ``entropy'' the quantity $S \equiv T_Xn_e^{-2/3}$, where $n_e$ is the electronic density and $T_X$ the deprojected temperature of the ICM \citep{ponman99}. In this paper, we use the definition of ``pseudo-entropy''  $s \equiv T*(EM)^{-1/3}$, where $T$ is the projected temperature and $EM$ is the emission measure, i.\,e.\, the normalization of the MEKAL model in XSPEC per units of area (see \citealt{rossetti06} for more details).}:
\begin{equation}
\label{def_sigma}
\sigma \equiv \frac{s_{IN}}{s_{OUT}}=\frac{T_{IN}}{T_{OUT}}\left(\frac{EM_{IN}}{EM_{OUT}}\right)^{-1/3}.
\end{equation}
The pseudo-entropy ratio is well correlated with the entropy ratio (see Appendix A) and therefore it is a useful and easy-to-calculate indicator of the variation of the physical three-dimensional entropy in the cores of galaxy clusters. \\
As in Paper I, we compare our pseudo-entropy classification with two alternative classification schemes, based on the core properties (Col.\, 6 in Table \ref{clusters}) and on the dynamical state (Col.\, 7 in Table \ref{clusters}). 
More specifically, in Col.\, 6 we divide clusters in three classes: cool core (CC), intermediate systems (INT) and non cool core (NCC), where CC  feature a prominent surface brightness peak and a temperature gradient, NCC possess neither of these properties, while INT show only one of these observational features. In Col.\, 7 we divide clusters in two classes: major mergers (MRG) and clusters showing no evidence of a major mergers (NOM). We consider evidence of a major merger the presence of cluster-wide diffuse radio emission, multi peaked velocity distribution of galaxies and significant irregularities on the surface brightness and temperature maps. The lack of these properties is not sufficient to state that an object is relaxed, this is why we prefer to consider these objects as ``no observed mergers''. We refer to Paper I for more details on the classification schemes and for the necessary references.\\
For the purposes of this paper, we are interested in separating clusters with a low entropy core (LEC) from the rest, this is why we have defined only one threshold to divide LEC from non-LEC objects. With our selected threshold ($s_{IN}/s_{OUT}=0.45$), we can state that all LEC clusters are known to be CC and do not show any evidence of merger (except A85, probably undergoing a merger in an early stage, which has not affected the observational properties of the core, see Paper I). Concerning non-LEC clusters, they are almost all non cool core systems, and many of them show significant indications of a major merger.  As already discussed, the fact that some of them do not show significant merger features does not mean that they are relaxed. 

\begin{table*}
\caption{Properties of the clusters in the sample: redshift (as derived from NED), ``mean'' temperature (i.\,e.\, $T_{OUT}$) and $R_{180}$. Col.\,5 lists the values of $s_{IN}/s_{OUT}$ we have derived, Col.\, 6 and Col.\, 7 list alternative classification schemes (see text). The first 14 entries are LEC clusters, while the remaining are non-LEC.}
\label{clusters}
\centering
\begin{tabular}{c c c c c c c}     
\hline\hline  
Name & Redshift & Temperature & $R_{180}$  & $\sigma$ & Core Cl. & Dyn.\, Cl. \\     
 & & (keV) & (Mpc) & & \\
\hline
A2199   &   0.0301   &   $4.101 \pm 0.025$   &    1.58 &     $ 0.4216  \pm    0.0033 $  &   CC & NOM \\
A496    &   0.0329   &   $4.990 \pm 0.038$   &    1.75 &     $ 0.3109  \pm   0.0028$ & CC & NOM$^*$ \\
2A0335+096 & 0.035   &   $3.649 \pm 0.009$   &    1.49 &     $ 0.2644  \pm   0.0007 $ & CC & NOM \\
A2052   &   0.035 &     $2.884  \pm    0.014$  &    1.33   &   $0.4274  \pm    0.0028$ &  CC & NOM \\ 
A4059    &   0.047   &   $3.968    \pm  0.028$     &  1.55    &   $0.4394   \pm   0.0044$  & CC & NOM \\
Hydra A  &   0.054    &   $3.413    \pm   0.024 $  &    1.43   &    $0.4138 \pm      0.0040 $ & CC & NOM \\
A85   &   0.055   &    $5.491 \pm      0.048$    &  1.82    &  $0.3857    \pm   0.0050$ &  CC & MRG \\ 
A1795  &     0.062   &    $5.416  \pm   0.026$   &    1.80  &  $0.3524  \pm     0.0023$ & CC & NOM \\
A3112   &    0.075  &    $ 4.319  \pm   0.033$   &    1.59   &    $0.3855 \pm     0.0038$ & CC & NOM \\
A2597    &  0.085  &  $ 3.491  \pm    0.018 $  &   1.43   &  $ 0.3568  \pm   0.0026$& CC & NOM \\
A478   &   0.088  &    $7.057   \pm    0.034$  &    2.03   &   $0.3706  \pm     0.0024$ & CC & NOM$^*$  \\
PKS0745-191  &    0.103    &  $7.765   \pm    0.12$  &    2.11    &  $0.2957    \pm    0.0056$& CC & NOM$^*$ \\
A2204    &   0.152  &     $7.412   \pm    0.092$   &    2.01   &    $0.2972  \pm     0.0046$  & CC & NOM \\
A2029    &  0.077  &   $ 6.173   \pm   0.056$  &    1.90   &   $0.4505 \pm   0.0057$ & CC & NOM \\
\hline
A4038   &   0.030    &  $2.957   \pm  0.015$   &  1.35   &   $0.5093    \pm  0.0037$ & INT & NOM\\
A576     &  0.039  &      $3.777   \pm    0.044 $  &     1.52   &    $0.6226   \pm   0.0135  $ & INT & MRG \\
A3571    &  0.039   &    $6.253  \pm     0.045 $   &    1.95 &     $0.6220  \pm    0.0078$ & INT & NOM\\
A119    &   0.044  &     $6.032  \pm     0.082$  &     1.92  &    $0.7621    \pm    0.0279$ & NCC & MRG\\
MKW3s   &   0.045  &    $3.325   \pm   0.018$  &    1.42  &    $0.4860   \pm   0.0039$ & CC & NOM \\
A1644    &   0.047  &      $4.156  \pm    0.056 $    &   1.59     &   $0.5318   \pm    0.0132 $ & INT & NOM\\
A3558     &  0.048    &   $5.221  \pm    0.025$   &    1.78    &   $0.6155  \pm   0.0057$  & INT & NOM \\
A3562     & 0.048  &    $4.344  \pm   0.032$   &   1.62    &  $0.6089  \pm    0.0083$ & NCC & MRG\\
Triangulum Australis   &    0.051   &    $9.224  \pm    0.095$   &    2.36  &    $0.6845   \pm   0.0164$ & NCC & NOM\\
A3158      & 0.060  &    $4.903   \pm   0.042 $  &    1.71     & $0.7488  \pm    0.0153$ & NCC & MRG\\
A399  &    0.072  &    $5.984  \pm   0.109$  &    1.88  &    $0.7086  \pm    0.0319$ & NCC & MRG\\
A401   &   0.074  &  $  7.268   \pm   0.084$ &    2.07  &   $ 0.7028   \pm    0.0212$ & NCC & MRG\\
A2255   &    0.081   &  $   6.197  \pm   0.154  $   &    1.91    &   $  0.8853   \pm    0.0714 $  & NCC & MRG\\
A1650   &    0.084   &  $  5.427 \pm    0.0539  $ &    1.78  &    $0.5323  \pm    0.0089$ &  INT & NOM \\
A1689     &  0.183    &  $ 8.612   \pm     0.102 $  &    2.13     &  $0.4724    \pm   0.0089$ &  INT & NOM\\
A2319   &   0.056   &   $8.962  \pm     0.091 $  &   2.32  &   $ 0.5643   \pm    0.0113$ & NCC & MRG\\
A3532   &    0.055 &   $  4.839    \pm   0.104 $ &     1.70  &     $0.7285   \pm    0.0336$ & INT & NOM$^*$\\
A3667    &   0.056  &   $  5.461 \pm      0.029 $  &    1.81  &  $   0.6378 \pm     0.0078$&  NCC & MRG$^*$\\
A754    &   0.054   &   $ 8.544   \pm   0.102  $  &   2.27   &  $  0.5289   \pm   0.0120$&  NCC & MRG \\
A3266   &    0.055 &   $  7.351   \pm   0.081 $  &    2.10   & $   0.6772   \pm   0.0170$&  NCC & MRG$^*$  \\
A2256   &    0.057  &   $  5.709  \pm    0.0763 $  &    1.85   &  $  0.6300   \pm   0.0197$&  NCC & MRG$^*$\\

\hline                  
\end{tabular}
\begin{list}{}{}
\item [Notes] 
$^*$ Properties and the references for the clusters not included in the sample in Paper I. A496: CC and no evidence for a major merger (\citealt{tamura01} and references therein). A478: CC and no evidence for a major merger (\citealt{sanderson05} and references therein). PKS0745-191: CC and no evidence for a major merger (\citealt{chen03} and references therein).
A3532: INT (moderate temperature drop) and no evidence for a major merger (the indications of optical substructures, \citealt{bardelli00}, are insufficient to claim a major merger). A3667: NCC and MRG (evidence of radio relics \citealt{rott97}, optical substructures \citealt{owers09}, and disturbed thermodynamical maps \citealt{briel04, vikh01b}). A3266: NCC and MRG (even if  extended radio emission has not been detected, \citealt{buote01}, there are evidences of optical substructures,  \citealt{quintana96}, and disturbed thermodynamical maps \citealt{finog06}). A2256: NCC and MRG (radio halo and relic, \citealt{kim99}, optical substructures, \citealt{berrington02}, and disturbed temperature maps, \citealt{sun02, bourdin08}).
\end{list}
\end{table*}

\section{Cool core remnants}
\label{CC_remnants}

\begin{figure}

   \centering
   \resizebox{\hsize}{!}{\includegraphics{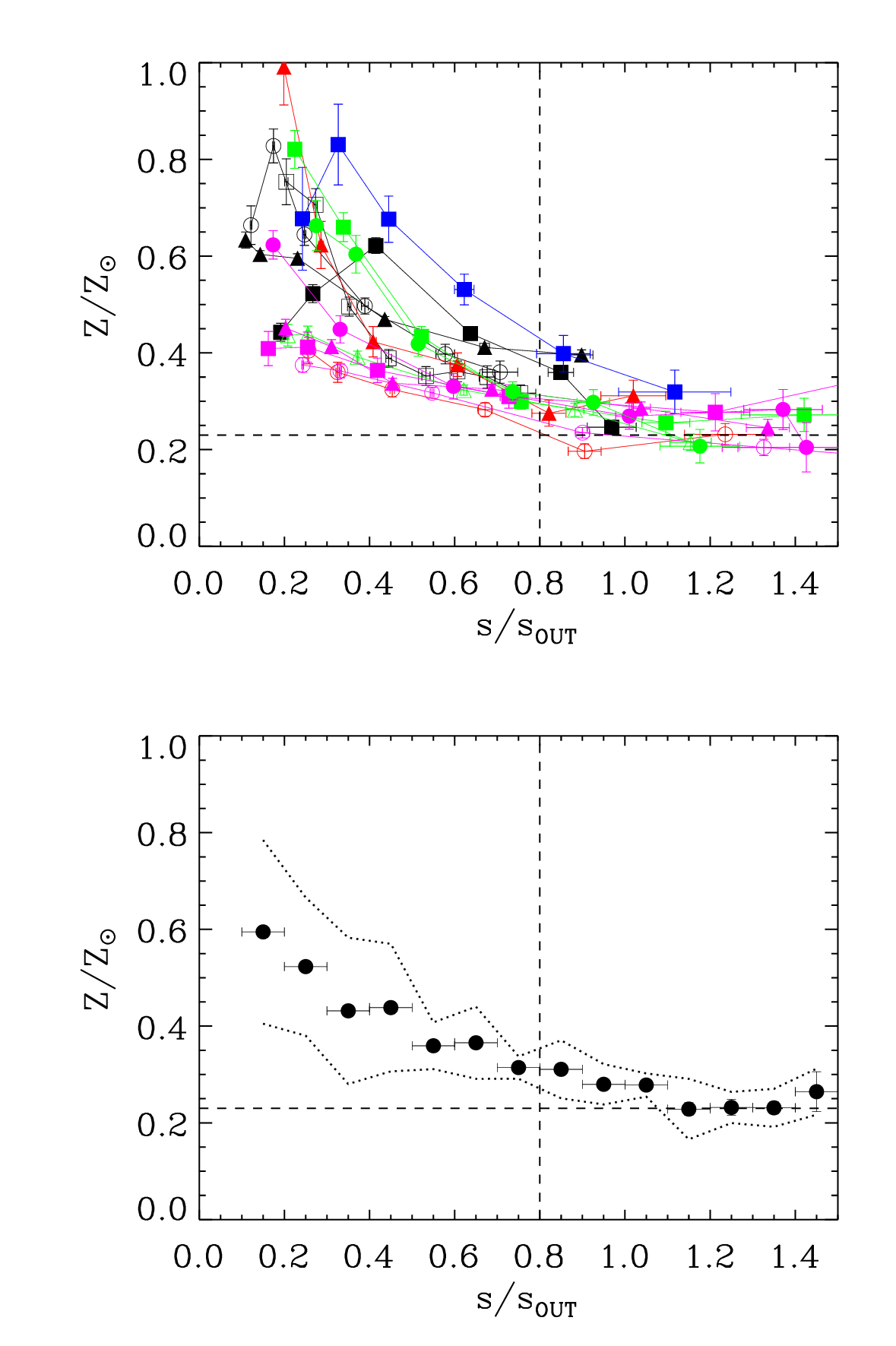}} 
\caption{\emph{Upper panel:} Metal abundance profiles of LEC clusters as a function of the pseudo-entropy ratio, in the regions of spectral extraction. The symbols and colors are as in Table \ref{tablist}. \emph{Lower panel:} Mean error weighted abundance profile for LEC clusters. Error bars show the (small) error on the average, while dotted lines show the $1\sigma$ scatter around the mean.
In both panels, the horizontal dashed line indicates the mean value of outer regions of galaxy clusters $Z=0.23Z_{\sun}$ \citep{leccardi08b} and the vertical dashed line indicates the selected threshold of $s/s_{OUT}$ under which all LEC cluster show a significant metal abundance excess.}
   \label{rat_Z_ps}
 \end{figure}

The central regions of LEC clusters in our sample are characterized by low-entropy (by definition), often accompanied by a temperature decrease and a surface brightness excess. These regions are also characterized by a high metal abundance: the mean iron abundance in the cores of LEC clusters is significantly larger than the typical value of  outer regions of galaxy clusters $Z=0.23Z_{\sun}$ \citep{leccardi08b}, even if with a $\simeq 20\%$ scatter \citep{degrandi09}.
It has been shown that the excess abundance is
consistent with being due to metals ejected from the BCG galaxy that is invariably found in these systems \citep{degrandi04}.
At present, we are not aware of any galaxy cluster with central regions characterized by  low entropy gas without a metal abundance excess. Therefore, there must be a close relation between entropy and metal abundance, and generally speaking, between the thermodynamics and the chemical properties of the ICM. An example of this relation can be seen in Fig.\,\ref{rat_Z_ps}, where, for each LEC cluster,   we plot the metal abundance radial profile as a function of the pseudo-entropy ratio. The pseudo entropy ratio is defined as in Equation \ref{def_sigma}, with the one difference that the temperature and emission measure of a given annulus replace $T_{IN}$ and $EM_{IN}$, i.\,e.\, $s(r)/s_{OUT}=T(r)/T_{OUT}*(EM(r)/EM_{OUT})^{-1/3}$. 
In the lower panel of Fig.\,\ref{rat_Z_ps}, we plot the mean error weighted abundance profile as a function of the pseudo-entropy ratio for LEC clusters and the one $\sigma$ scatter around the average of the values. Low entropy ICM is characterized by a a significant metal abundance excess, although the scatter is quite large for $s/s_{out}<0.4$. The plots show that all LEC clusters show a significant metal abundance excess with respect to the outer mean value $Z=0.23Z_{\sun}$ \citep{leccardi08b} in regions characterized by a pseudo entropy ratio smaller than $0.8$ (corresponding to physical radii $r\lesssim 0.07 R_{180}$). \\
In Paper I, we found that  some clusters without a low-entropy core presented an unusually high metal abundance in their IN region (as already discussed such high abundances are typical of LEC). We concluded that the most likely
explanation for the high central abundance of these anomalous non LEC systems was that at some time in the past they hosted a cool core (i.e. low-entropy gas and metal abundance excess) that  was subsequently heated up. Indeed, a heating event that does not completely disrupt a cool core will most likely leave behind a region characterized by a high metalicity and by an entropy that, albeit not as low as the one found in the central regions of LEC systems, will be lower
than that found in other regions of the cluster. We emphasize that metals are reliable markers of the ICM, in the sense that, once they have polluted a given region of a cluster, the time scale over which they will diffuse is comparable to the Hubble time \citep{sar88}. Thus metals trace the ICM where they have been injected.
 The presence of these regions has also been predicted by the simulations of \citet{motl04}, who suggested that even if the observational cool core signature may disappear, cool gas (and we may add ``low-entropy and metal rich gas'') remains in the merging systems at all times. \\
Since in LEC clusters a metal abundance excess is invariably associated to ICM with a low pseudo-entropy, we have systematically selected the regions with the lowest pseudo-entropy ratio in non-LEC systems, with the aim of finding regions possibly characterized by a high metalicity.
In practice, for the 21 non-LEC clusters of our sample, we have prepared two-dimensional pseudo-entropy ratio maps, by dividing the pseudo-entropy maps for the pseudo-entropy in the OUT region derived with  spectral analysis. In these maps we have identified regions characterized by a $s/s_{OUT}<0.8$ (as shown in Fig.\,\ref{rat_Z_ps} all LEC clusters show a metal abundance excess in regions with entropy ratio $<0.8$), and we have extracted and analyzed spectra in these regions (as described in Sect. \ref{prof}) to investigate their metalicity. Pseudo-entropy ratio maps and figures of the selected regions are provided in Appendix B.\\
\begin{figure}
   \centering
   \resizebox{\hsize}{!}{\includegraphics{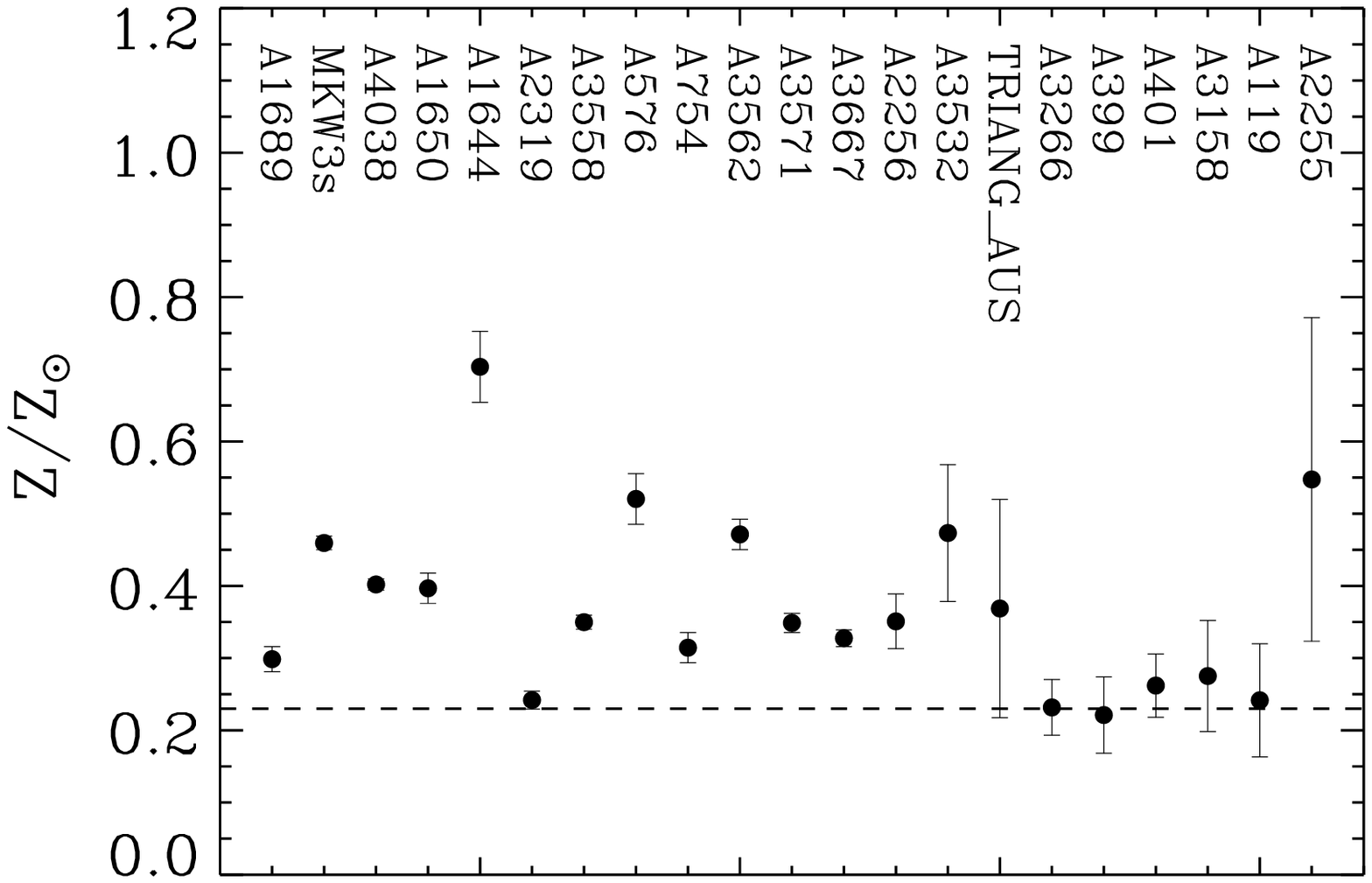}}\\
   \resizebox{\hsize}{!}{\includegraphics{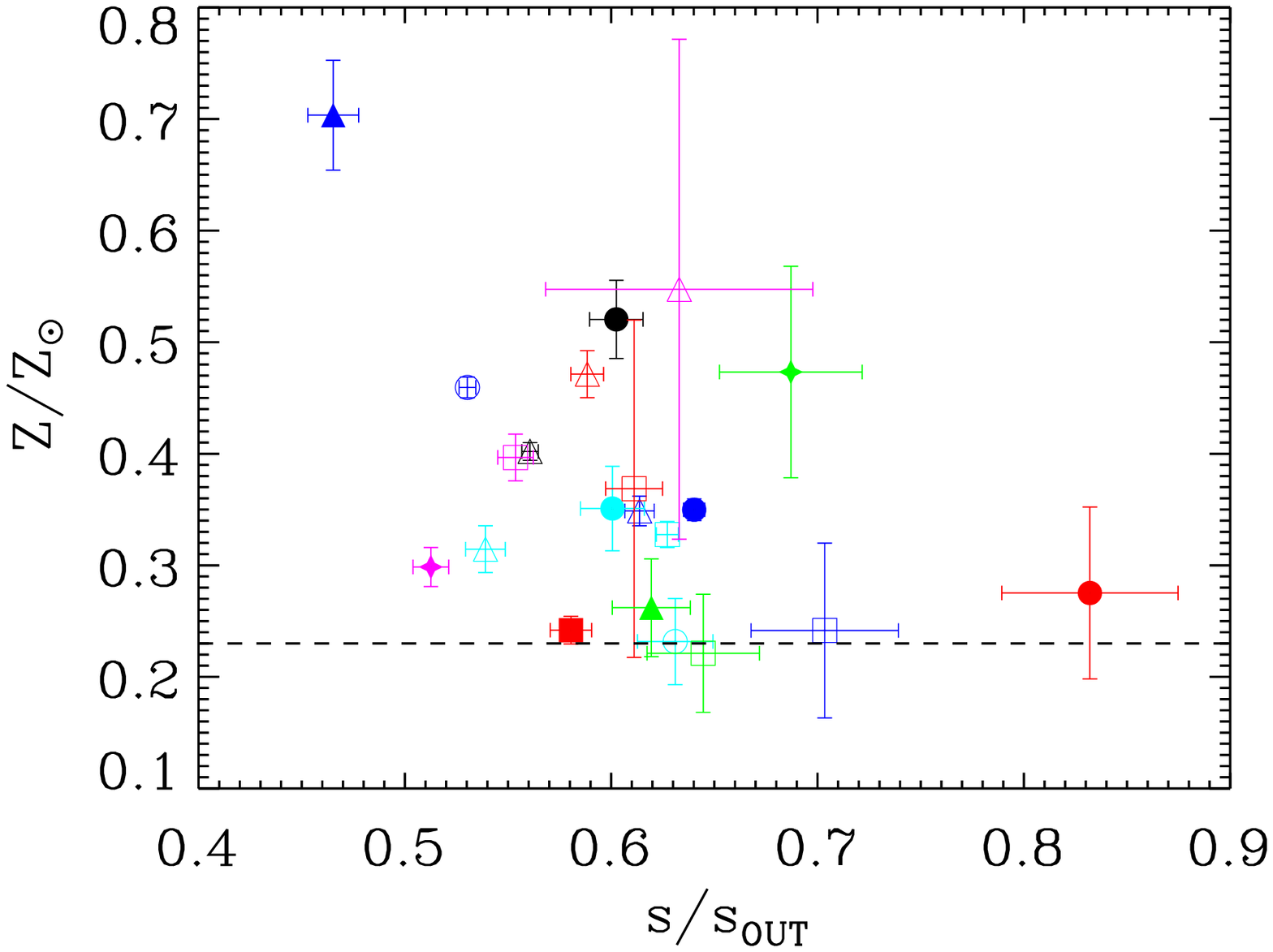}}
   \caption{Metal abundance in the regions selected for spectral analysis for the 21 non-LEC objects in the sample (upper panel) and as a function of the pseudo-entropy ratio in the regions (lower panel). The horizontal line represents the reference value for the metalicity of of outer regions of galaxy clusters $Z=0.23Z_{\sun}$ \citep{leccardi08b}. Symbols and colors are as in Table \ref{tablist}.}
   \label{met_reg}
 \end{figure}
In all the objects (21) of our subsample of non-LEC clusters, we find regions characterized by an entropy ratio smaller than $0.8$. In Fig. \ref{met_reg}, we plot the metal abundance in these regions for all the clusters (results are reported in Table \ref{met_reg_tab}).  12 objects (namely: A1689,  MKW3s, A4038, A1650, A1644, A3558, A576, A754, A3562,  A3571,  A3667 and A2256) show a significant excess ($>3\sigma$ c.l.) with respect to the reference value $Z=0.23Z_{\sun}$ \citep{leccardi08b}. 
In the remaining 9 clusters, the metal abundance is statistically consistent with the reference value. Amongst these objects, we will consider as clusters without metal abundance excess those where the metalicity is measured with a good statistical accuracy, more specifically the 6 clusters where  $Z>0.4Z_{\sun}$ can be excluded at more than $2\sigma$ confidence level (A2319, A3266, A399, A401, A119 and A3158). We will no more consider in this paper the three objects (A3532, Triangulum Australis and A2255) where the error bars on the metal abundance are too large to discriminate between the two classes (the observations of these clusters are the shortest of our sample, see Table 1).\\
The most likely interpretation is that the low entropy high Z regions found in many of our non-LEC systems are what remains of a cool core after a heating event. We dub these structures ``cool core remnants''\footnote{The term ``remnant of cool cores'' has been recently used by \citet{fusco09} to describe structures similar to the ones we detect albeit without information on the metal abundance.}, a more detailed discussion of how they formed and of alternative scenarios is provided in Section \ref{discussion}. \\
In the lower panel of Fig. \ref{met_reg} we plot the metal abundance as a function of the pseudo entropy ratio, $s/s_{OUT}$ where $s$ is the pseudo-entropy in the region used for spectral analysis.
It is worth noting that the clusters with no indication of a significant metal abundance excess are those where the selected regions show the highest $s/s_{OUT}$, with the notable exception of A2319 (red filled square) \\
To better characterize the systems in our sample, we have compared the entropy maps with the galaxy distribution and more specifically the low-entropy regions with the position of the BCG (Col.\, 3 in Table \ref{met_reg_tab} and Figures in Appendix B). The name and positions of the BCGs are derived from the literature (\citealt{coziol09, lin04, gudehus91, postman95}). 
In 9/12 clusters with ``CC remnants'' (namely  A1650, MKW3s, A4038, A1644, A3558, A3562, A3571, A1689 and A576) the BCGs are located in the low-entropy high metal abundance regions.
In A3667 and A2256, the positions of the BCGs do not coincide with the ``CC remnant'', which is moreover not obviously associated to any galaxy concentration (a more detailed study of the galaxy distribution in A3667 by \citealt{owers09} found essentially the same results). 
 In A754 the BCG is located $\sim 700$ kpc west of the selected regions but in the ``CC remnant'' we find the elliptical galaxy 2MASSX J09091923-0941591 (see Figure in Appendix B), which is the brightest member of the secondary peak in the galaxy distribution \citep{fabricant86}. The position of this secondary galaxy concentration, possibly associated to the CC remnant, coincides also with a clump in the dark matter distribution, as derived with gravitational lensing \citep{okabe08}.
For the non-LEC objects where we do not observe a significant metalicity excess, we find that in 3/6 systems (A119, A401 and A3158) the low entropy regions are not obviously associated to the BCGs. In the remaining 3 clusters (A399, A3266, and A2319) the BCGs are located in the low-entropy regions where we do not find evidence of a metal excess.\\
The main reason for which we compared the position of CC remnants with the galaxy distribution is that galaxies, and more specifically BCGs, are responsible of the metal abundance excess observed in LEC clusters \citep{degrandi04}. However the galaxy distribution may also provide information on the shape of the potential well of the clusters. Indeed, in relaxed clusters BCGs are located at the bottom of the potential well and their positions coincide with the centroid of the X-ray isophotes and with the brightness peak of their host clusters. We have compared the position of CC remnants with the position of the X-ray peak, the centroid of X-ray isophotes and the position of the BCGs (Figures in Appendix B). In most cases (namely A1650, MKW3s, A4038, A1644, A1689, A576, A3562,  and A3571),  these three points lie within the low entropy regions even if they do not necessarily coincide with one another. However, in the remaining clusters one or more of these indicators is significantly offset from the CC remnant. A754 and A2256 have a very disturbed morphology and the X-ray centroid does not coincide with the CC remnant or with the BCG. In A3667 the centroid roughly coincides with the brightness ``peak'' and the BCG, but the CC remnant is offset by almost 500 kpc. It is apparent that in these last three cases, the low entropy metal rich regions are not in equilibrium within the cluster's potential well and therefore we are observing a rapidly evolving situation.

\begin{table}
\caption{Properties of low entropy regions. In col.\,2 we list the metal abundance in solar units, while in col.\,3 we write ``Y'' if the BCG is found in the low-entropy region and ``N'' if it is elsewhere. 
Clusters in italic are ``AGN CC-remnant''  while clusters in bold type are ``merger CC-remnant'' (Sect.\, \ref{discussion} for these definitions).}
\label{met_reg_tab}
\centering                          
\begin{tabular}{c c c}
\hline\hline  
Cluster & Metal Abundance (solar) & BCG  \\
\hline
\it{A1650}   &   $0.397   \pm  0.021$ & Y \\
\it{MKW3s}   &   $0.459   \pm   0.009$ & Y \\
\it{A4038} $^\mathrm{a}$   &   $0.402    \pm  0.008$ & Y  \\
\it{A1644} $^\mathrm{a}$  &   $0.703   \pm  0.049$ & Y \\
\bf{A1689}   &   $0.298   \pm   0.017$ &  Y  \\
\bf{A3558}   &   $0.350  \pm    0.010$ & Y \\
\bf{A576}    &  $0.520   \pm    0.035$ & Y \\
\bf{A754}    &  $0.314     \pm  0.021$ & N$^{\mathrm{b}}$\\
\bf{A3562}   &   $0.471   \pm   0.021$ & Y \\
\bf{A3571}   &   $0.349    \pm  0.013$ & Y \\
\bf{A3667}   &   $0.327    \pm 0.012$ & N \\
\bf{A2256}   &   $0.351   \pm   0.038$ & N \\
A2319   &   $0.242   \pm    0.012$ & Y \\
A3266   &   $0.232   \pm  0.039$& Y \\
A399    &  $0.221   \pm  0.053$ & Y \\
A401    &  $0.262   \pm   0.044$ & N \\
A3158   &  $ 0.275   \pm  0.077$ & N \\
A119    &  $0.242    \pm  0.078$ & N \\
A2255 $^\mathrm{c}$   &  $0.547   \pm   0.224$ & \\
A3532  $^\mathrm{c}$  &   $0.473    \pm  0.095$ & \\
Triangulum Australis $^\mathrm{c}$  &     $0.369  \pm    0.151$ &  \\
\hline                                   
\end{tabular}
\begin{list}{}{}
\item[Notes:] $^{\mathrm{a}}$ Even if the central entropy of this cluster is within reach of AGN outbursts, it also show some indications of on-going interaction and cannot be unambiguously associated to the ``AGN CC-remnant class'' (see text). $^{\mathrm{b}}$ The BCG 2MASSX J09083238-0937470 is not associated to the CC-remnant, which however coincides with the position of the elliptical galaxy 2MASSX J09091923-0941591. $^\mathrm{c}$ Due to the large indetermination on the metal abundance estimate, we could not unambiguously detect or exclude the presence of a metal abundance excess in this object. \\
\end{list}

\end{table}

\section{Discussion}
\label{discussion} 
In Section \ref{CC_remnants}, we have shown that 12 out of 21 non-LEC objects of our sample, host regions characterized by low entropy and high metal abundance, with respect to mean values in the outer ICM. Such conditions are usually verified in the cores of relaxed LEC clusters, albeit with stronger gradients (i.e. the entropy of the ICM reaches smaller values).
 We have dubbed these features ``cool core remnants'', since we have interpreted them as what remains of a cool core after a ``heating event''. \\
One may argue that these structures have evolved similarly but independentely of LEC systems and that there is no need to trace them back to known structures, such as LEC. In the ``primordial'' scenario, where non-LEC clusters started off with different early condition and are now slowly cooling to become LEC, low entropy and metal rich region could be interpreted as ``progenitors'' of cool cores, rather than remnants. However this scenario does not explain why in some cases these structures are not in equilibrium within the potential well of their host clusters and are not associated to the BCG or to a giant elliptical galaxy. On the contrary, these issues are naturally addressed in  the ``evolutionary'' scenario if we consider the effects of mergers both on the ICM and on the galaxy distribution (Sec.\, \ref{sub_merger}).\\
Moreover we should pay attention to the fact that most of these structures are found in clusters undergoing major mergers, i.\,e.\, in rapidly evolving objects. It is therefore necessary to trace these transient structures back to an equilibrium state from which they recently evolved.\\
Another possible objection is that the low-entropy structures may not be  embedded in the ICM, but rather in groups along the line of sight of the clusters. However, in the cases where the cool core remnants are associated to the BCGs or to a galaxy concentration (see Sect.\, \ref{CC_remnants}), the redshifts of the galaxies are consistent with the mean readshift of the clusters. In the cases where there is no galaxy concentration obviously associated to the low entropy regions, these features would have to be interpreted as ``gas-only'' groups superposed on the line of sight.\\
In the interpretation of the low-entropy regions as CC remnants, the ``heating event'' is responsible for smoothing the entropy gradient of the cores of LEC clusters. We have identified two possible mechanisms: interaction with the central AGN and mergers. \\

\subsection{CC remnants and central AGNs}
\label{sub_agn}
\begin{figure}
   \centering
   \resizebox{\hsize}{!}{\includegraphics{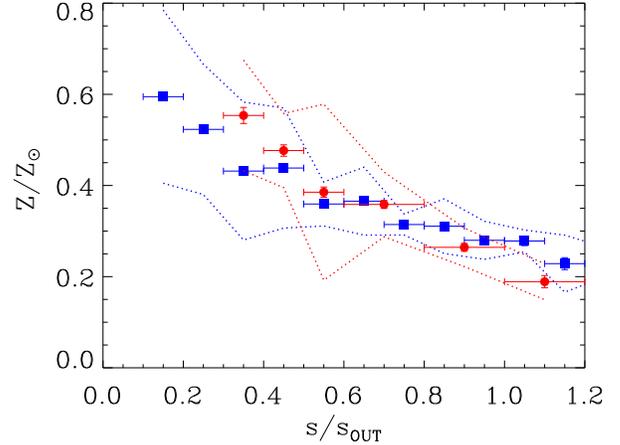}}\\
   \caption{Error weighted metal abundance profile as a function of the pseudo entropy ratios for LEC clusters (blue squares) and for the four clusters with low central entropy (red circles). Dotted lines show the one $\sigma$ scatter of the data.}
   \label{LEC_AGN}
 \end{figure}

Interactions between the central AGN and the ICM have been observed with \chandra and \xmmn in many LEC clusters and these interactions are now considered as the principal mechanism preventing the formation of cooling flows (\citealt{peterson06} and references therein). In some cases, powerful AGN outbursts may increase significantly the entropy content of some galaxy clusters, transforming a low entropy core into a non-LEC object. This is possible for systems where the core entropy has been raised to a relatively small value $s_0\simeq (30-50)\, \rm{keV}
\, \rm{cm}^{2}$ \citep{voit05}.
In our subsample of non-LEC clusters, according to \emph{Chandra} measurements \citep{cava09}, only A1650, A4038, MKW3s and A1644 have core entropy smaller than  $50\, \rm{keV} \, \rm{cm}^{2}$ and therefore are within the reach of powerful AGNs. AGN heating is the most plausible explanation for increasing the entropy in the cores of A1650 (\citealt{donahue05, voit05}) and for MKW3s (radio lobes have been observed in this cluster, \citealt{giacintucci06}). We will consider these clusters as ``AGN CC-remnants''\\.
The remaining two non-LEC clusters with a relatively low central core entropy, A4038 and A1644, are controversial objects, showing some  indications of ongoing interactions, contrary to A1650 and MKW3s which look relaxed at all wavelengths (see Table 2 in Paper I). 
A4038 does not have a large temperature drop at the center but it has a low central cooling time. This is why some authors classify it as a cool core \citep{peres98} and others as a non cool core \citep{sanderson06}. In the optical, \citet{burgett04} found some indications of substructure on large scales. 
A1644 is probably an advanced off-axis merger with two subclumps clearly visible in the X-ray image and some indications of on-going sloshing of the cool core of the main subclump \citep{reiprich04}.
The interpretation of A4038 and A1644 is not straightforward, since merging events could have  contributed to the heating of the cores (see Sect.\,\ref{sub_merger}). \\
In AGN CC remnants, the low entropy regions coincide with the centers of the clusters and with the positions of the BCGs as expected. 
In Fig.\, \ref{LEC_AGN} we compare the error weighted mean metal abundance profile of LEC clusters with the same profile for non-LEC clusters with $s_0<50\, \rm{keV} \, \rm{cm}^{2}$. In the innermost bins (lowest entropy ratios), the mean profile of this class of clusters is significantly larger than in LEC clusters, although the scatter is large. This is consistent with a scenario where a localized heating in the inner regions of clusters has increased the entropy ratio without substantially modifying the metal abundance, resulting in ICM with enhanced metalicity for its entropy.\\ 

\subsection{CC remnants and mergers}
\label{sub_merger}
As shown in Sect.\, \ref{sub_agn}, 8 out of 12 clusters featuring a metal abundance excess have core entropies too large to be produced even by the most powerful AGN outbursts. Therefore, we need another mechanism which could have produced the CC-remnants that we observe today.
The fact that most of these clusters (namely A754, A3667, A2256, A3562 and A576) show strong indications of a major merger suggests that merging could provide the necessary heating to produce these features.
The indications are not as strong for the remaining three clusters (A1689, A3558 and A3571; in Table \ref{clusters} they are classified as ``intermediate'' and ``no observed merger''). However, as shown in the notes reported below, they all possess some  peculiar features, which suggest that they are undergoing some kind of interaction. It is therefore natural to define this class of 8 objects as ``merger CC remnants''.

\begin{center}
 \emph{Notes on A1689, A3558 and A3571}
\end{center}
\begin{description}
 \item[\bf{A1689}]
This cluster has been considered for a long time by X-ray astronomers as an example of a relaxed cluster, due to its spherical shape (Figure in Appendix B) and very peaked surface brightness profile (e.\,g. \, \citealt{peres98}). However, analysis of the distribution of the velocities of the galaxies showed two distinct peaks, suggesting a line of sight superposition \citep{girardi01}. More recently, an analysis of \xmmn data by \citet{andersson04} showed asymmetric temperature and redshift distribution which, combined with optical data, could suggest an on-going merger. 
\item[\bf{A3558}]
\citet{rossetti06} classify this cluster as an intermediate object, since it has some characteristics of CC but it also shows some indications of an on-going merger, like asymmetry in the temperature and entropy distribution and substructures in the 3-D galaxy distributions \citep{bardelli98b}. Due to the very peculiar environment in which it is located (the core of the Shapley Supercluster), these features may be explained as the results of multiple minor mergers or of a past off-axis major merger with A3562.
\item[\bf{A3571}] 
This cluster has a regular morphology in X-rays (Figure in Appendix B), but we do not observe a temperature decrease in the center. This is why it is often classified as a ``non cool core'' object \citep{ohara06, sanderson06}. On the basis of its radio and optical properties, \citet{venturi02} suggest that this cluster is a late stage merger.
\end{description}

We recall here the main results concerning the position of merger CC remnants with respect to the centroid of X-ray emission and to the galaxy distribution (Sect.\, \ref{CC_remnants} and Table \ref{met_reg_tab}). In 5/8 clusters (namely  A3558, A3562, A3571, A1689, A576) the BCGs and the X-ray centroid are located in the low-entropy high metal abundance regions.
 In A754 the CC remnant is associated to a secondary peak in the galaxy distribution (with a giant elliptical galaxy), but it is also offset from the position of the large scale X-ray centroid.
In A3667 and A2256 the CC remnants are not obviously associated to any galaxy concentration and they are not found at the centers of their host clusters (see figures in Appendix B).
These different results are not in contrast with our interpretation since the effect of the merging processes both on the collisional ICM and on the non-collisional dark matter and galaxies depends on many parameters such as the merger state, the mass-ratios and the impact parameter. Therefore, while in A3667 and A2256 the low entropy and high metal abundance gas may have completely decoupled from its galaxy concentration and possibly dark matter halo (such as in the ``bullett cluster'', \citealt{clowe06}), in A3558, A3562, A3571, A1689, A576 and A754 this gas has  not been completely displaced from its gravitational potential well. Another possibility is that, in the latter systems, we are currently observing a late stage of a merger, where the low entropy high metal abundance gas is slowly moving back to the center of the cluster.\\
Two main mechanisms can contribute to the formation of a CC remnant during a merger event: shock heating and mixing. Mergers are expected to drive moderately supersonic shock waves in the ICM, which, being irreversible changes, increase the entropy of the system. If one of the merging subcluster had a CC before the merging event, shock heating may significantly increase the entropy of the core but in some cases this heating may not be sufficient to bring the entropy to the large values observed in some non-LEC clusters. With a very simplified calculation based on Rainkine-Hugoniot jump conditions, we estimate that shocks with Mach  number $1-3$ can increase the mean entropy in the cores only of a factor up to $1.8$. 
Mergers also drive motions in the ICM in which low entropy metal-rich gas may be put in contact with high-entropy metal poor ICM. The mean entropy of the resulting mixed ICM will be lower than ambient entropy (but higher than typical core entropy) and its metal content could be larger than in the ambient gas. However, since there is a significant spread in the metal abundance values of cool cores, mixing may also efficiently disrupt small abundance gradients.  We recall here that metals are frozen in the ICM and therefore mixing with metal poor ICM is the only possible way of reducing the metal abundance. \\
\begin{figure}
    \centering
   \resizebox{\hsize}{!}{\includegraphics{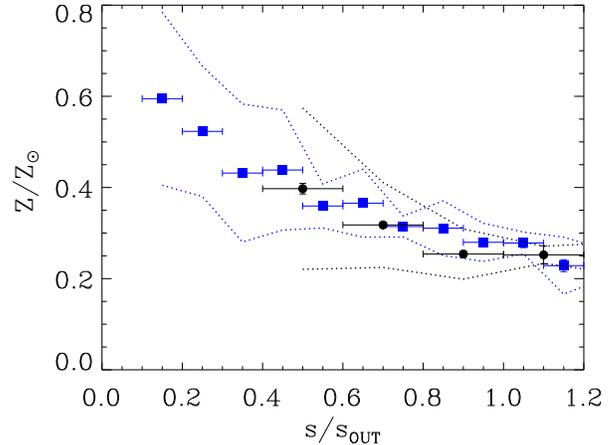}}\\
  \caption{Error weighted metal abundance profile as a function of the pseudo entropy ratios for LEC clusters (blue squares) and for the clusters with merger CC remnants  (red circles). Dotted lines show the one $\sigma$ scatter of the data.}
  \label{LEC_MER}	
\end{figure}
 In Fig.\, \ref{LEC_MER} we compare the mean metal abundance profile for clusters with merger CC remnants with that of LEC clusters, similarly to Fig.\, \ref{LEC_AGN} for AGN CC remnants. In this case the mean profiles are consistent but the scatter in clusters with CC remnants is larger. Indeed, we did not expect the same results as in Fig.\, \ref{LEC_AGN} (i.\,e.\, gas with enhanced metalicity for its entropy) because, contrary to AGN heating, merger heating does not affect only the central regions of clusters and therefore the entropy ratios are modified in a non-trivial way. Moreover we cannot neglect the effect of gas mixing, both on the entropy and on the metal abundance.

\subsection{Non LEC clusters without metal abundance excess}
Of the 21 objects of our sample which are classified as non LEC, 6 features regions with pseudo-entropy ratios smaller than $0.8$ but without a significant metal abundance excess ($Z<0.4Z_{\sun}$ at more than $2\sigma$ confidence level). We will discuss in this section some of the properties of these systems.\\ 
As shown in Table \ref{clusters}, we have strong indications of ongoing mergers for 5 out of 6 objects, namely A2319, A399, A401, A119 and A3266.
The remaining cluster, A3158, shows some indications of ongoing interactions based mainly on the galaxy distribution \citep{johnston08}. 
These clusters do not have a CC remnant, and this could mean that they have never developed a cool core, as in scenario proposed by \citet{mccarthy08}. However we should note that while in this ``primordial'' scenario, NCC may indifferently be relaxed or non relaxed objects, in most objects of this class (if not all) we have found indications of on-going interactions. \\
 The fact that most of these clusters show indications of on-going interactions, could suggest that they may have had a low-entropy core that has been completely erased during the merger event. As shown in many simulation works, the effects of a merger on the structure of the ICM depends on many parameters, such as the mass ratio, the impact parameter and the structure of the interacting subclusters. Therefore it is not surprizing that in some cases the merger can efficiently destroy the low entropy core, while in others it leaves a CC remnant.\\
We recall here than in 3  clusters (A399, A3266 and A2319) the BCGs are located in the low-entropy region where we do not find evidence of a metal excess (see Table \ref{met_reg_tab}). The interpretation of this class of clusters is an interesting task: we speculate that either the LEC state from which they evolved was characterized by a metal abundance profile showing only a moderate excess or, alternatively, that the mixing may have been more effective in these clusters than in those where we observe CC remnants.
 We recall that metalicity profiles of LEC clusters show a large scatter: as can be seen also in Fig.\, \ref{rat_Z_ps}, while the central metal abundance of some clusters can reach almost solar values, in others it reaches only $0.4\, Z_{\sun}$. Assuming a constant degree of mixing in all cluster, it is clear that it would be easier to detect a metal excess in regions where gas with  $Z\sim\, Z_{\sun}$ has mixed with the ambient gas, than in regions where mixing has occurred between ICM with $Z\simeq 0.4\, Z_{\sun} $  and ICM with $Z\simeq 0.2\, Z_{\sun} $. Therefore systems like A399, A3266 and A2319 could be remnants of ``metal poor CC'', where ``poor'' means that the metal abundance is lower with respect to other CC objects.
Alternatively, these could be systems where metals have been mixed more efficiently than in others where an abundance excess is detected. \\
It is interesting to note that the fraction of objects where the BCG is not associated to the low-entropy regions is higher in clusters showing no $Z$ excess (3/6) than in merger CC remnants (2/8), even if we should note that this difference is not statistically significant. As shown in many simulations and in the case of the ``bullett cluster'' \citep{clowe06}, in violent major mergers the collisionless galaxy population may completely decouple from the collisional ICM. It is natural to assume that these violent mergers are also more effective in mixing the gas and therefore in completely erasing metal abundance gradients. In this scenario, the three clusters showing no metal abundance excess and no BCG associated, namely A119, A3158 and A401,  should be those that have undergone the most destructive interactions. However this speculation needs further verification with a larger sample of clusters.
\subsection{Metal abundance and entropy distribution}
\begin{figure}
   \centering
   \resizebox{\hsize}{!}{\includegraphics{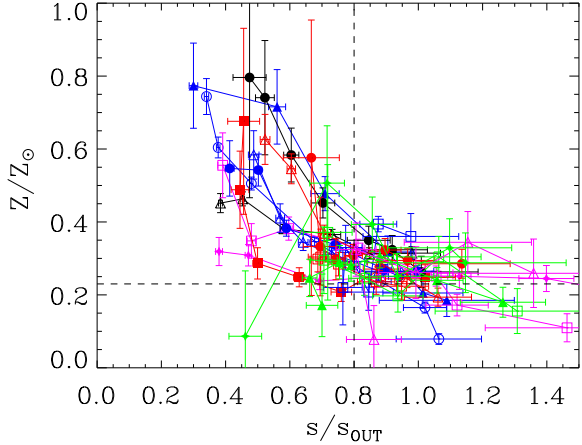}}\\
   \caption{Metal abundance profiles of non-LEC clusters as a function of the pseudo-entropy ratio, in the regions of spectral extraction. The symbols and colors are as in Table \ref{tablist}. The dashed horizontal line indicates the mean value of outer regions of galaxy clusters $Z=0.23Z_{\sun}$, \citep{leccardi08b}, while the vertical dashed lines marks the threshold pseudo entropy ratio used to select regions for spectral analysis.}
   \label{Z_ps_nonLEC}
 \end{figure}

One may speculate whether our selection based on the pseudo-entropy ratio $<0.8$ is effective in identifying all the regions with a large metalicity. 
Indeed, a possible counter-example to our choice of using an entropy ratio threshold for the selection of regions for spectral analysis would be the observation of a large metal abundance associated to high entropy gas and generally speaking  the lack of anti-correlation between metal abundance and pseudo-entropy ratio.\\
In Fig.\, \ref{rat_Z_ps} we have shown that the metal abundance excess in LEC clusters is typically found in regions with $s/s_{OUT}<0.8$. In a few cases, an excess is detected also in regions with $s/s_{OUT}\sim 1$  but not at larger values.
Moreover, the metal abundance is always anti-correlated with the entropy.\\
We have performed the same test on the subsample of non-LEC clusters, using the results of spectral analysis in radial annuli centered on the surface brightness peak (Fig.\, \ref{Z_ps_nonLEC}). We do not include in this figures the four highly disturbed clusters for which we have not performed spectral analysis in radial annuli. 
It is interesting to note that we do not find a single region where a significant metal excess is associated to a pseudo entropy ratio larger than 1.
If a metal abundance excess is observed it is invariably associated to ICM with a low pseudo-entropy ratio.\\

\subsection{Cooling times}
The results presented in this paper support the ``evolutionary'' scenario of the CC-NCC dichotomy: most non-LEC clusters have likely ``evolved'' from a LEC state.
It is therefore natural to ask ourselves if this is a ``one way trip'', i.\,e.\, if once a cluster has undergone a merger it remains a non-LEC cluster for the rest of its life  or it has the possibility of reforming a low-entropy core, as in the old cyclic vision of cooling flows-mergers.\\
\begin{figure}
   \centering
   \resizebox{\hsize}{!}{\includegraphics{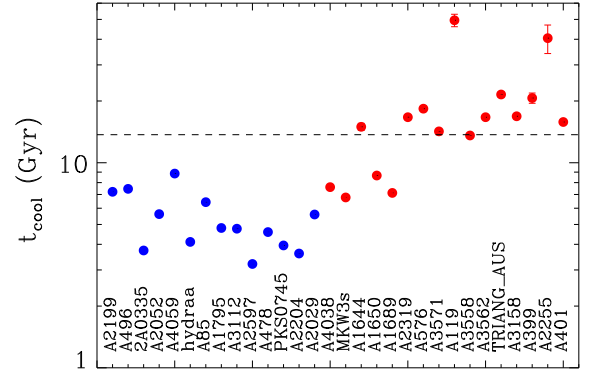}}\\
   \caption{Cooling time in the inner $0.05 \, R_{180}$ for the clusters of the sample for which we have performed radial analysis and de-projection. Horizontal dashed line is the Hubble time.}
   \label{tcool}
 \end{figure}
In Fig.\, \ref{tcool} we plot the cooling time in a central bin of radius $0.05\, R_{180}$ for the clusters of the initial sample for which we have performed radial analysis and de-projection (i.\,e.\, excluding A3266, A2256, A3667 and A754).
The cooling time is generally larger than the Hubble time for non-LEC clusters, including those of most clusters with CC-remnant (A576, A3571, A3562, A3558, we do not have this 3-D information for A3667, A754 and A2256). Therefore these clusters would appear to be unable to develop a new low-entropy core.\\ 
The cooling time shown in Fig.\, \ref{tcool} is a mean value calculated in large region. If the mixing of the ICM is not completely effective, clumps of cool gas may reside in these regions and in these clumps the cooling will be much more effective. In this case, we should consider the measured cooling time only as an upper limit and it is possible that the cooler gas could contribute to the formation of a low entropy core on time scales shorter than the Hubble time. Any future observation of these clumps of cool gas, will be useful to test the possibility of a ``return journey'' to the LEC state. 

\section{Summary and Conclusions}
In this paper, we have presented a systematic two-dimensional analysis of the entropy and metal abundance of a sample of nearby clusters observed with \xmmn. 
\begin{itemize}
 \item We have analyzed the observations of a sample of 35 clusters. Following the prescription in Paper I, we have classified 14 of them as low-entropy cores. In the remaining 21 non-LEC objects we have performed a systematical analysis of the entropy maps  and selected regions with pseudo-entropy ratio lower than $ 0.8$, for a proper spectral analysis.  All the clusters in our subsample host regions with these characteristics.
\item In most non-LEC objects (12/21) the low entropy regions are characterized by a significant metal abundance excess with respect to outer regions of galaxy clusters. For 6 of the remaining clusters, we can significantly exclude an abundance larger than $0.4Z_{\sun}$, while for 3 objects the error bars are too large to discriminate.
We have interpreted the low-entropy high metal abundance regions that we found in 12 clusters as what remains of a cool core after a heating event, and we have dubbed them ``CC remnants''.
\item For two clusters, A1650 and MKW3s, the most likely heating mechanism is AGN feedback. As shown by \citet{voit05}, clusters with core entropy $\lesssim 50\, \rm{keV}\, \rm{cm}^{2}$  are within the reach of powerful AGN outbursts.
\item In 8/12 clusters with CC remnant, the core entropy is too large to be produced by giant AGN outbursts. In all these clusters we have found indication of on-going interactions and 5 of them are probably undergoing a major merger. Therefore the most likely heating mechanism for this class of objects is merging, which can shock heat and mix the ICM. In most of these clusters, the CC remnant is associated to the BCG or to a giant elliptical galaxy which could be responsible for the production of the metal excess during the LEC phase. In A3667 and A2256 the merging may have completely decoupled the low-entropy metal rich gas from its associated galaxy concentration. In A754, A2256 and A3667 the position of the CC remnant does not
coincide with the centroid of the X-ray emission, indicating that the low entropy metal rich ICM is not in equilibrium within the potential well of the host clusters. 
\item In 6 objects out of 21, we have not found a significant metal excess in the low entropy regions. Since most of these clusters show indications of  major mergers, we have interpreted them as clusters where the metal abundance gradient has been erased by the mixing following mergers.
\end{itemize}
The results presented in this paper strengthens those shown in Paper I. Indeed, in Paper I we found a metal abundance excess in a small fraction of non-LEC clusters. Thanks to the two-dimensional analysis presented in this paper, we have found a metal abundance excess in most objects. We conclude that most non-LEC objects have spent part of their life as low-entropy cores. After a heating event the LEC signature disappears but the ``CC-like'' ICM (i.\,e.\, characterized by large metal abundance and low entropy, albeit not as low as in LEC) remains in the systems and can be identified in two-dimensional entropy maps.
In the framework of the alternative ``primordial'' scenario the low entropy metal rich regions could be interpreted as ``progenitors'' of cool cores, that are now evolving (slowly cooling and increasing their metalicity) to become LEC.  However this scenario does not explain why most of these regions lie in dynamically active objects and at least three of them are found in a configuration of non-equilibrium, offset from the center of the clusters. Conversely,  this issue is naturally addressed if they are the results of a merger event in the evolutionary scenario.\\
The results summarized above strongly support scenarios where cluster core properties are not fixed ``ab initio'' but evolve across cosmic time. 

\begin{acknowledgements}
We acknowledge useful discussions with Fabio Gastaldello and Sabrina De Grandi.
This research has made use of the NASA/IPAC Extragalactic Database (NED), of the High Energy Astrophysics Science Archive (HEASARC), of the \xmmn Science Archive (XSA), of the Digitized Sky Surveys (DSS) and of the Archive of \chandra Cluster Entropy Profiles Tables (ACCEPT). 
\end{acknowledgements}

\bibliographystyle{aa}
\bibliography{refs}

\begin{thebibliography}{75}
\expandafter\ifx\csname natexlab\endcsname\relax\def\natexlab#1{#1}\fi

\bibitem[{Anders \& Grevesse(1989)}]{anders89}
Anders, E. \& Grevesse, N. 1989, \gca, 53, 197

\bibitem[{{Andersson} \& {Madejski}(2004)}]{andersson04}
{Andersson}, K.~E. \& {Madejski}, G.~M. 2004, \apj, 607, 190

\bibitem[{{Arnaud} {et~al.}(2005){Arnaud}, {Pointecouteau}, \&
  {Pratt}}]{arnaud05}
{Arnaud}, M., {Pointecouteau}, E., \& {Pratt}, G.~W. 2005, \aap, 441, 893

\bibitem[{{Ascasibar} \& {Markevitch}(2006)}]{ascas06}
{Ascasibar}, Y. \& {Markevitch}, M. 2006, \apj, 650, 102

\bibitem[{{Bardelli} {et~al.}(1998){Bardelli}, {Pisani}, {Ramella}, {Zucca}, \&
  {Zamorani}}]{bardelli98b}
{Bardelli}, S., {Pisani}, A., {Ramella}, M., {Zucca}, E., \& {Zamorani}, G.
  1998, \mnras, 300, 589

\bibitem[{{Bardelli} {et~al.}(2000){Bardelli}, {Zucca}, {Zamorani},
  {Moscardini}, \& {Scaramella}}]{bardelli00}
{Bardelli}, S., {Zucca}, E., {Zamorani}, G., {Moscardini}, L., \& {Scaramella},
  R. 2000, \mnras, 312, 540

\bibitem[{{Bauer} {et~al.}(2005){Bauer}, {Fabian}, {Sanders}, {Allen}, \&
  {Johnstone}}]{bauer05}
{Bauer}, F.~E., {Fabian}, A.~C., {Sanders}, J.~S., {Allen}, S.~W., \&
  {Johnstone}, R.~M. 2005, \mnras, 359, 1481

\bibitem[{{Berrington} {et~al.}(2002){Berrington}, {Lugger}, \&
  {Cohn}}]{berrington02}
{Berrington}, R.~C., {Lugger}, P.~M., \& {Cohn}, H.~N. 2002, \aj, 123, 2261

\bibitem[{{Bourdin} \& {Mazzotta}(2008)}]{bourdin08}
{Bourdin}, H. \& {Mazzotta}, P. 2008, \aap, 479, 307

\bibitem[{{Briel} {et~al.}(2004){Briel}, {Finoguenov}, \& {Henry}}]{briel04}
{Briel}, U.~G., {Finoguenov}, A., \& {Henry}, J.~P. 2004, \aap, 426, 1

\bibitem[{{Brunetti} {et~al.}(2009){Brunetti}, {Cassano}, {Dolag}, \&
  {Setti}}]{brunetti09}
{Brunetti}, G., {Cassano}, R., {Dolag}, K., \& {Setti}, G. 2009, to appear in
  \aap, astro-ph/0909.2343

\bibitem[{{Buote}(2001)}]{buote01}
{Buote}, D.~A. 2001, \apjl, 553, L15

\bibitem[{{Burgett} {et~al.}(2004){Burgett}, {Vick}, {Davis}, {Colless}, {De
  Propris}, {Baldry}, {Baugh}, {Bland-Hawthorn}, {Bridges}, {Cannon}, {Cole},
  {Collins}, {Couch}, {Cross}, {Dalton}, {Driver}, {Efstathiou}, {Ellis},
  {Frenk}, {Glazebrook}, {Hawkins}, {Jackson}, {Lahav}, {Lewis}, {Lumsden},
  {Maddox}, {Madgwick}, {Norberg}, {Peacock}, {Percival}, {Peterson},
  {Sutherland}, \& {Taylor}}]{burgett04}
{Burgett}, W.~S., {Vick}, M.~M., {Davis}, D.~S., {et~al.} 2004, \mnras, 352,
  605

\bibitem[{{Burns} {et~al.}(1997){Burns}, {Loken}, {Gomez}, {Rizza}, {Bliton},
  \& {Ledlow}}]{burns97}
{Burns}, J.~O., {Loken}, C., {Gomez}, P., {et~al.} 1997, in Astronomical
  Society of the Pacific Conference Series, Vol. 115, Galactic Cluster Cooling
  Flows, ed. N.~{Soker}, 21--+

\bibitem[{{Cappellari} \& {Copin}(2003)}]{cappellari}
{Cappellari}, M. \& {Copin}, Y. 2003, \mnras, 342, 345

\bibitem[{{Cavagnolo} {et~al.}(2009){Cavagnolo}, {Donahue}, {Voit}, \&
  {Sun}}]{cava09}
{Cavagnolo}, K.~W., {Donahue}, M., {Voit}, G.~M., \& {Sun}, M. 2009, \apjs,
  182, 12

\bibitem[{{Chen} {et~al.}(2003){Chen}, {Ikebe}, \& {B{\"o}hringer}}]{chen03}
{Chen}, Y., {Ikebe}, Y., \& {B{\"o}hringer}, H. 2003, \aap, 407, 41

\bibitem[{{Clowe} {et~al.}(2006){Clowe}, {Brada{\v c}}, {Gonzalez},
  {Markevitch}, {Randall}, {Jones}, \& {Zaritsky}}]{clowe06}
{Clowe}, D., {Brada{\v c}}, M., {Gonzalez}, A.~H., {et~al.} 2006, \apjl, 648,
  L109

\bibitem[{{Coziol} {et~al.}(2009){Coziol}, {Andernach}, {Caretta},
  {Alamo-Mart{\'{\i}}nez}, \& {Tago}}]{coziol09}
{Coziol}, R., {Andernach}, H., {Caretta}, C.~A., {Alamo-Mart{\'{\i}}nez},
  K.~A., \& {Tago}, E. 2009, \aj, 137, 4795

\bibitem[{{De Grandi} {et~al.}(2004){De Grandi}, {Ettori}, {Longhetti}, \&
  {Molendi}}]{degrandi04}
{De Grandi}, S., {Ettori}, S., {Longhetti}, M., \& {Molendi}, S. 2004, \aap,
  419, 7

\bibitem[{{De Grandi} \& {Molendi}(2009)}]{degrandi09}
{De Grandi}, S. \& {Molendi}, S. 2009, to appear in \aap, astro-ph/0909.1224

\bibitem[{{De Luca} \& {Molendi}(2004)}]{deluca03}
{De Luca}, A. \& {Molendi}, S. 2004, \aap, 419, 837

\bibitem[{{Diehl} \& {Statler}(2006)}]{diehl06}
{Diehl}, S. \& {Statler}, T.~S. 2006, \mnras, 368, 497

\bibitem[{{Donahue} {et~al.}(2005){Donahue}, {Voit}, {O'Dea}, {Baum}, \&
  {Sparks}}]{donahue05}
{Donahue}, M., {Voit}, G.~M., {O'Dea}, C.~P., {Baum}, S.~A., \& {Sparks}, W.~B.
  2005, \apjl, 630, L13

\bibitem[{{Dunn} \& {Fabian}(2008)}]{dunn08}
{Dunn}, R.~J.~H. \& {Fabian}, A.~C. 2008, \mnras, 385, 757

\bibitem[{{Dunn} {et~al.}(2005){Dunn}, {Fabian}, \& {Taylor}}]{dunn05}
{Dunn}, R.~J.~H., {Fabian}, A.~C., \& {Taylor}, G.~B. 2005, \mnras, 364, 1343

\bibitem[{{Edge} {et~al.}(1990){Edge}, {Stewart}, {Fabian}, \&
  {Arnaud}}]{edge90}
{Edge}, A.~C., {Stewart}, G.~C., {Fabian}, A.~C., \& {Arnaud}, K.~A. 1990,
  \mnras, 245, 559

\bibitem[{{Ettori} {et~al.}(2002){Ettori}, {De Grandi}, \&
  {Molendi}}]{ettori02}
{Ettori}, S., {De Grandi}, S., \& {Molendi}, S. 2002, \aap, 391, 841

\bibitem[{{Fabricant} {et~al.}(1986){Fabricant}, {Beers}, {Geller},
  {Gorenstein}, {Huchra}, \& {Kurtz}}]{fabricant86}
{Fabricant}, D., {Beers}, T.~C., {Geller}, M.~J., {et~al.} 1986, \apj, 308, 530

\bibitem[{{Ferrari} {et~al.}(2008){Ferrari}, {Govoni}, {Schindler}, {Bykov}, \&
  {Rephaeli}}]{ferrari08}
{Ferrari}, C., {Govoni}, F., {Schindler}, S., {Bykov}, A.~M., \& {Rephaeli}, Y.
  2008, Space Science Reviews, 134, 93

\bibitem[{{Finoguenov} {et~al.}(2006){Finoguenov}, {Henriksen}, {Miniati},
  {Briel}, \& {Jones}}]{finog06}
{Finoguenov}, A., {Henriksen}, M.~J., {Miniati}, F., {Briel}, U.~G., \&
  {Jones}, C. 2006, \apj, 643, 790

\bibitem[{{Fusco-Femiano} {et~al.}(2009){Fusco-Femiano}, {Cavaliere}, \&
  {Lapi}}]{fusco09}
{Fusco-Femiano}, R., {Cavaliere}, A., \& {Lapi}, A. 2009, to appear in \apj,
  astro-ph/0909.2943

\bibitem[{{Giacintucci} {et~al.}(2006){Giacintucci}, {Mazzotta}, {Brunetti},
  {Venturi}, \& {Bardelli}}]{giacintucci06}
{Giacintucci}, S., {Mazzotta}, P., {Brunetti}, G., {Venturi}, T., \&
  {Bardelli}, S. 2006, Astronomische Nachrichten, 327, 573

\bibitem[{{Girardi} \& {Mezzetti}(2001)}]{girardi01}
{Girardi}, M. \& {Mezzetti}, M. 2001, \apj, 548, 79

\bibitem[{{G{\'o}mez} {et~al.}(2002){G{\'o}mez}, {Loken}, {Roettiger}, \&
  {Burns}}]{gomez02}
{G{\'o}mez}, P.~L., {Loken}, C., {Roettiger}, K., \& {Burns}, J.~O. 2002, \apj,
  569, 122

\bibitem[{{Gudehus} \& {Hegyi}(1991)}]{gudehus91}
{Gudehus}, D.~H. \& {Hegyi}, D.~J. 1991, \aj, 101, 18

\bibitem[{{Henry} {et~al.}(2004){Henry}, {Finoguenov}, \& {Briel}}]{henry04}
{Henry}, J.~P., {Finoguenov}, A., \& {Briel}, U.~G. 2004, \apj, 615, 181

\bibitem[{{Johnston-Hollitt} {et~al.}(2008){Johnston-Hollitt}, {Sato}, {Gill},
  {Fleenor}, \& {Brick}}]{johnston08}
{Johnston-Hollitt}, M., {Sato}, M., {Gill}, J.~A., {Fleenor}, M.~C., \&
  {Brick}, A.-M. 2008, \mnras, 390, 289

\bibitem[{{Kim}(1999)}]{kim99}
{Kim}, K.-T. 1999, Journal of Korean Astronomical Society, 32, 75

\bibitem[{{Leccardi} \& {Molendi}(2008{\natexlab{a}})}]{leccardi08b}
{Leccardi}, A. \& {Molendi}, S. 2008{\natexlab{a}}, \aap, 487, 461

\bibitem[{{Leccardi} \& {Molendi}(2008{\natexlab{b}})}]{leccardi08a}
{Leccardi}, A. \& {Molendi}, S. 2008{\natexlab{b}}, \aap, 486, 359

\bibitem[{{Leccardi} {et~al.}(2009){Leccardi}, {Molendi}, \& {Rossetti}}]{pap1}
{Leccardi}, A., {Molendi}, S., \& {Rossetti}, M. 2009, to appear in \aap,
  astro-ph/0910.4894 (Paper I)

\bibitem[{{Lin} \& {Mohr}(2004)}]{lin04}
{Lin}, Y.-T. \& {Mohr}, J.~J. 2004, \apj, 617, 879

\bibitem[{{Maccacaro} {et~al.}(1988){Maccacaro}, {Gioia}, {Wolter}, {Zamorani},
  \& {Stocke}}]{mac88}
{Maccacaro}, T., {Gioia}, I.~M., {Wolter}, A., {Zamorani}, G., \& {Stocke},
  J.~T. 1988, \apj, 326, 680

\bibitem[{{McCarthy} {et~al.}(2008){McCarthy}, {Babul}, {Bower}, \&
  {Balogh}}]{mccarthy08}
{McCarthy}, I.~G., {Babul}, A., {Bower}, R.~G., \& {Balogh}, M.~L. 2008,
  \mnras, 386, 1309

\bibitem[{{McCarthy} {et~al.}(2004){McCarthy}, {Balogh}, {Babul}, {Poole}, \&
  {Horner}}]{mccarthy04}
{McCarthy}, I.~G., {Balogh}, M.~L., {Babul}, A., {Poole}, G.~B., \& {Horner},
  D.~J. 2004, \apj, 613, 811

\bibitem[{Molendi \& Pizzolato(2001)}]{molepizzo01}
Molendi, S. \& Pizzolato, F. 2001, \apj, 560, 194

\bibitem[{{Motl} {et~al.}(2004){Motl}, {Burns}, {Loken}, {Norman}, \&
  {Bryan}}]{motl04}
{Motl}, P.~M., {Burns}, J.~O., {Loken}, C., {Norman}, M.~L., \& {Bryan}, G.
  2004, \apj, 606, 635

\bibitem[{{O'Hara} {et~al.}(2006){O'Hara}, {Mohr}, {Bialek}, \&
  {Evrard}}]{ohara06}
{O'Hara}, T.~B., {Mohr}, J.~J., {Bialek}, J.~J., \& {Evrard}, A.~E. 2006, \apj,
  639, 64

\bibitem[{{Okabe} \& {Umetsu}(2008)}]{okabe08}
{Okabe}, N. \& {Umetsu}, K. 2008, \pasj, 60, 345

\bibitem[{{Owers} {et~al.}(2009){Owers}, {Couch}, \& {Nulsen}}]{owers09}
{Owers}, M.~S., {Couch}, W.~J., \& {Nulsen}, P.~E.~J. 2009, \apj, 693, 901

\bibitem[{{Peres} {et~al.}(1998){Peres}, {Fabian}, {Edge}, {Allen},
  {Johnstone}, \& {White}}]{peres98}
{Peres}, C.~B., {Fabian}, A.~C., {Edge}, A.~C., {et~al.} 1998, \mnras, 298, 416

\bibitem[{{Peterson} \& {Fabian}(2006)}]{peterson06}
{Peterson}, J.~R. \& {Fabian}, A.~C. 2006, \physrep, 427, 1

\bibitem[{Peterson {et~al.}(2001)Peterson, Paerels, Kaastra, Arnaud, Reiprich,
  Fabian, Mushotzky, Jernigan, \& Sakelliou}]{peterson01}
Peterson, J.~R., Paerels, F. B.~S., Kaastra, J.~S., {et~al.} 2001, \aap, 365, L
  104

\bibitem[{{Ponman} {et~al.}(1999){Ponman}, {Cannon}, \& {Navarro}}]{ponman99}
{Ponman}, T.~J., {Cannon}, D.~B., \& {Navarro}, J.~F. 1999, \nat, 397, 135

\bibitem[{{Poole} {et~al.}(2008){Poole}, {Babul}, {McCarthy}, {Sanderson}, \&
  {Fardal}}]{poole08}
{Poole}, G.~B., {Babul}, A., {McCarthy}, I.~G., {Sanderson}, A.~J.~R., \&
  {Fardal}, M.~A. 2008, \mnras, 391, 1163

\bibitem[{{Postman} \& {Lauer}(1995)}]{postman95}
{Postman}, M. \& {Lauer}, T.~R. 1995, \apj, 440, 28

\bibitem[{{Quintana} {et~al.}(1996){Quintana}, {Ramirez}, \&
  {Way}}]{quintana96}
{Quintana}, H., {Ramirez}, A., \& {Way}, M.~J. 1996, \aj, 112, 36

\bibitem[{{Reiprich} {et~al.}(2004){Reiprich}, {Sarazin}, {Kempner}, \&
  {Tittley}}]{reiprich04}
{Reiprich}, T.~H., {Sarazin}, C.~L., {Kempner}, J.~C., \& {Tittley}, E. 2004,
  \apj, 608, 179

\bibitem[{{Ritchie} \& {Thomas}(2002)}]{ritchie02}
{Ritchie}, B.~W. \& {Thomas}, P.~A. 2002, \mnras, 329, 675

\bibitem[{Rossetti(2006)}]{rossettith}
Rossetti, M. 2006, PhD thesis, Univ. of Milano

\bibitem[{{Rossetti} {et~al.}(2007){Rossetti}, {Ghizzardi}, {Molendi}, \&
  {Finoguenov}}]{rossetti06}
{Rossetti}, M., {Ghizzardi}, S., {Molendi}, S., \& {Finoguenov}, A. 2007, \aap,
  463, 839

\bibitem[{{Rottgering} {et~al.}(1997){Rottgering}, {Wieringa}, {Hunstead}, \&
  {Ekers}}]{rott97}
{Rottgering}, H.~J.~A., {Wieringa}, M.~H., {Hunstead}, R.~W., \& {Ekers}, R.~D.
  1997, \mnras, 290, 577

\bibitem[{{Sanderson} {et~al.}(2009{\natexlab{a}}){Sanderson}, {Edge}, \&
  {Smith}}]{sanderson09b}
{Sanderson}, A.~J.~R., {Edge}, A.~C., \& {Smith}, G.~P. 2009{\natexlab{a}},
  \mnras, 398, 1698

\bibitem[{{Sanderson} {et~al.}(2005){Sanderson}, {Finoguenov}, \&
  {Mohr}}]{sanderson05}
{Sanderson}, A.~J.~R., {Finoguenov}, A., \& {Mohr}, J.~J. 2005, \apj, 630, 191

\bibitem[{{Sanderson} {et~al.}(2009{\natexlab{b}}){Sanderson}, {O'Sullivan}, \&
  {Ponman}}]{sanderson09}
{Sanderson}, A.~J.~R., {O'Sullivan}, E., \& {Ponman}, T.~J. 2009{\natexlab{b}},
  \mnras, 428

\bibitem[{{Sanderson} {et~al.}(2006){Sanderson}, {Ponman}, \&
  {O'Sullivan}}]{sanderson06}
{Sanderson}, A.~J.~R., {Ponman}, T.~J., \& {O'Sullivan}, E. 2006, \mnras, 372,
  1496

\bibitem[{{Sarazin}(1988)}]{sar88}
{Sarazin}, C.~L. 1988, {X-ray emission from clusters of galaxies} (Cambridge
  Astrophysics Series, Cambridge: Cambridge University Press, 1988)

\bibitem[{{Sun} {et~al.}(2002){Sun}, {Murray}, {Markevitch}, \&
  {Vikhlinin}}]{sun02}
{Sun}, M., {Murray}, S.~S., {Markevitch}, M., \& {Vikhlinin}, A. 2002, \apj,
  565, 867

\bibitem[{Tamura {et~al.}(2001)Tamura, Bleeker, Kaastra, Ferrigno, \&
  Molendi}]{tamura01}
Tamura, T., Bleeker, A., Kaastra, J., Ferrigno, C., \& Molendi, S. 2001, \aap,
  379, 107

\bibitem[{{Venturi} {et~al.}(2002){Venturi}, {Bardelli}, {Zagaria}, {Prandoni},
  \& {Morganti}}]{venturi02}
{Venturi}, T., {Bardelli}, S., {Zagaria}, M., {Prandoni}, I., \& {Morganti}, R.
  2002, \aap, 385, 39

\bibitem[{{Venturi} {et~al.}(2008){Venturi}, {Giacintucci}, {Dallacasa},
  {Cassano}, {Brunetti}, {Bardelli}, \& {Setti}}]{venturi08}
{Venturi}, T., {Giacintucci}, S., {Dallacasa}, D., {et~al.} 2008, \aap, 484,
  327

\bibitem[{{Vikhlinin} {et~al.}(2007){Vikhlinin}, {Burenin}, {Forman}, {Jones},
  {Hornstrup}, {Murray}, \& {Quintana}}]{vikh_garching}
{Vikhlinin}, A., {Burenin}, R., {Forman}, W.~R., {et~al.} 2007, in Heating
  versus Cooling in Galaxies and Clusters of Galaxies, ed. H.~{B{\"o}hringer},
  G.~W. {Pratt}, A.~{Finoguenov}, \& P.~{Schuecker}, 48--+

\bibitem[{{Vikhlinin} {et~al.}(2001){Vikhlinin}, {Markevitch}, \&
  {Murray}}]{vikh01b}
{Vikhlinin}, A., {Markevitch}, M., \& {Murray}, S.~S. 2001, \apj, 551, 160

\bibitem[{{Voit} \& {Donahue}(2005)}]{voit05}
{Voit}, G.~M. \& {Donahue}, M. 2005, \apj, 634, 955

\end{thebibliography}

\begin{appendix} 
\section{Relation between Pseudo-Entropy and Entropy ratios}
\begin{figure}
   \centering
   \resizebox{\hsize}{!}{\includegraphics{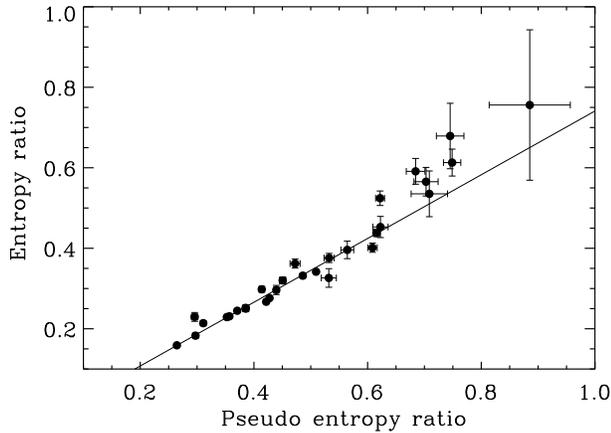}}\\
   \caption{Entropy ratio versus Pseudo-entropy ratio}
   \label{test_entro}
 \end{figure}
We have used the entropy profiles for the 31 clusters in our sample for which we have performed radial analysis, to test whether the pseudo-entropy ratios (defined in Paper I) are good indicators of the behavior of the true three dimensional entropy ($S\equiv T_{X}/n_e^{2/3}$). 
We have therefore calculated the entropy ratio $S_{IN}/S_{OUT}$, starting from the three-dimensional entropy profiles obtained with the de-projected temperature and density profiles. We show in Fig. \ref{test_entro} that there is a strong correlation between the ``true'' entropy ratio and the ``pseudo'' entropy ratios. We  fitted the points in Fig. \ref{test_entro} with a linear model taking into account the errors on both coordinates (with the IDL Astrolib procedure FITEXY.PRO\footnote{http://idlastro.gsfc.nasa.gov/ftp/pro/math/fitexy.pro}) and we found
\begin{equation}
\frac{S_{IN}}{S_{OUT}}=-0.051\pm0.004+(0.792 \pm 0.011)*\frac{s_{IN}}{s_{OUT}}.
\label{eq_a1}
\end{equation}
As expected, the entropy ratios are smaller than pseudo-entropy ratios, since projection effects smooth out gradients.\\
In order to quantify the scatter around the best fit relation, we have applied to the residuals a maximum likelihood algorithm that postulates a parent distribution described by a mean and an intrinsic dispersion \citep{mac88}. 
To do this we have calculated an ``effective error'' on the y coordinate, which takes into account the errors on both coordinates, defined as: $\sigma_{eff}^2=\sigma_y^2+b^2\sigma_x^2$, where $b$ is the best-fit value for the slope of the linear relation between the two data sets (in this case $b=0.792$, Eq.\, \ref{eq_a1}). The mean value of the residual distribution is consistent with zero ($0.021$), with an intrinsic dispersion $\sigma=0.043$.
\end{appendix}
\begin{appendix}
\section{Figures of individual clusters}
Large figures, available at:\\
http://www.iasf-milano.inaf.it/~rossetti/public/CCR/rossetti.pdf

\end{appendix}
\newpage

\longtab{1}{
\begin{longtable}{c c c c}
\caption{\label{tablist} Observations analyzed for the clusters in the sample. In case of multiple observations, the one indicated with $^*$ (usually the longest) is the one used for spectral analysis. The exposure time in the third column is the sum of the exposure times of the three instruments after SP cleaning. In Col.\, 4 we list the symbols and colours used to mark each cluster in the figures of the present paper.}\\
\hline \hline
Cluster & Obs ID & Exposure time (ks) & Symbol \\*
\hline
\endfirsthead
\caption{continued.}\\
\hline\hline
Cluster & Obs ID & Exposure time (ks) & Symbol\\
\hline
\endhead
\hline
\endfoot
\hline
\endlastfoot
A4038 &	0204460101 &	78.1 & Black open triangle\\	
A2199 &  0008030201$^*$	& 42.8 & Black open square\\*
  &  0008030301 & 11.2 \\*
  &  0008030601 & 11.3 \\
A496 &	0135120201 & 47.2 & Black open circle \\		
2A0335+096 &	 0147800201 &	230.8  & Black filled triangle\\	
A2052 &	0109920101 &	85.1 & Black filled square\\	
A576 & 0205070401$^*$ &	42.7 & Black filled circle\\*
  & 0205070301 & 28.1 \\
A3571 &	 0086950201 &	43.5 & Blue open triangle\\ 	
A119 &	0012440101 &	54.1 & Blue open square\\	
MKW3s &	0109930101 &	99.2 & Blue open circle \\
A1644 &	0010420201 &	42.0 & Blue filled triangle\\	
A4059 & 0109950101$^*$  &	32.4 & Blue filled square\\*
  &  0109950201 & 64.9 \\
A3558 &	0107260101 &	126.4 & Blue filled circle\\	
A3562 & 0105261301$^*$  &  116.6 & Red open triangle\\*  
  & 0105261501 & 45.0  \\*
  & 0105261601 & 52.5 \\*
  & 0105261701 & 57.7 \\*
  & 0105261801  & 23.6 \\
Triangulum Australis &  0093620101$^*$	&  27.3 & Red open square\\*
  & 0093620201 & 8.3 \\*
  & 0093620301 & 31.8\\
Hydra A & 0109980301 &	52.2 &  Red open circle\\	
A754 & 0136740101$^*$ &	40.6 & Light blue open square	\\*
   & 0136740201 & 14.8 \\*
  & 0112950401 & 31.8 \\*
 & 0112950301  & 34.8 \\
A3266 &  0105260901$^*$	 &  66.5 & Light blue open circle \\*
 & 0105262001 & 13.7 \\*
 & 0105262101 & 16.2 \\*
 & 0105261101 & 33.1 \\*
 & 0105262201 & 9.1 \\*
 & 0105261001 & 5.8 \\*
 & 0105260701 & 56.4 \\*
 & 0105260801 & 56.7 \\*
 & 0105262501 & 17.3 \\
 A85 &  0065140101	&	34.6  & Red filled triangle \\	
 A3532 & 0030140301 & 27.7 & Green filled star\\ 
A3667 &  0206850101$^*$	&	162.9  & Light blue open triangle \\*
  & 0105260101 & 15.4 \\*
  & 0105260601 & 62.5 \\*
  & 0105260401 & 45.6 \\*
  & 0105260301 & 47.2 \\*
  & 0105260501 & 38.6 \\*
  & 0105260201 & 31.8 \\
A2319 &  0302150101$^{\rm{a}} $	& 27.6$^{\rm{b}}$ & Red filled square \\*
  & 0302150201$^{\rm{a} }$ & 28.2$^{\rm{b}}$  \\ 
A2256 &  0141380201$^*$	& 33.1	& Light blue filled circle\\*
 & 0112950601 & 25.3 \\*
 & 0112951501 & 28.1 \\*
 & 0112951601 & 30.9 \\*
 & 0141380101 & 24.4 \\
A3158 & 0300210201$^*$	 &	33.2$^{\rm{b}}$	& Red filled circle \\*
  & 0300211301 & 14.0$^{\rm{b}}$	\\
A1795 &	0097820101 & 97.3 & Green open triangle\\	
A399 &	 0112260101 &	27.9 & Green open square\\	
A401 &	 0112260301 &	34.7 & Green filled triangle\\	
A3112 &	0105660101 &	64.6 & Green filled square\\	
A2029 &	0111270201 &	30.8 & Green filled circle\\	
A2255 &	0112260801 &	25.1 & Pink open triangle\\	
A1650 &	0093200101 &	75.0 & Pink open square\\	
A2597 &	0147330101&	144.3 & Pink open circle\\	
A478 & 	0109880101 &	136.7 & Pink filled triangle\\	
PKS0745-191 & 0105870101 &	34.3 & Pink filled square\\	
A2204 & 0112230301$^*$	&	51.2 & Pink filled circle	\\*
  & 0306490101 & 17.6$^{\rm{b}}$ \\*
  & 0306490201 & 30.9$^{\rm{b}}$ \\*
  & 0306490301 & 22.7$^{\rm{b}}$ \\*
  & 0306490401 & 29.7$^{\rm{b}}$ \\
A1689 &	0093030101 &	106.7 & Pink filled star\\
\end{longtable}
\begin{list}{}{}
\item[Notes:] $^{\rm{a}}$ We used both observations of A2319 to increase the statistics. $^{\rm{b}}$ Since these observations have been performed after degradation of CCD 6 in MOS1 in 2005, we did not consider the  
MOS1 detector and the exposure time given in the table is just MOS2 $+$ pn.
\end{list}
}

\end{document}